\documentclass[preprint,showpacs,preprintnumbers,amsmath,amssymb]{revtex4}
\usepackage{graphicx}% Include figure files
\usepackage{dcolumn}% Align table columns on decimal point
\usepackage{bm}% bold math
\usepackage{amssymb}
\usepackage{amsmath}
\usepackage{latexsym}

\begin{document}

%\draft
\title{%Hall and Shubnikov-De Haas Effects: Extension of the Drude model
Galvano- and thermo-magnetic effects at low and high temperatures within   non-Markovian quantum Langevin approach
}
\author{
I.B. Abdurakhmanov$^{1}$,
G.G. Adamian$^{2}$,  N.V. Antonenko$^{2}$, and Z. Kanokov$^{2,3,4}$
}
\affiliation{
$^{1}$Curtin Institute for Computation, Department of Physics, Astronomy and Medical Radiation Sciences, Curtin University,  Perth, WA 6845, Australia\\
$^{2}$Joint Institute for
Nuclear Research, 141980 Dubna, Russia\\
$^{3}$National University, 700174 Tashkent,
Uzbekistan\\
$^{4}$Institute of Nuclear Physics,
702132 Tashkent, Uzbekistan
%\\
%$^{4}$Institut f\"ur Theoretische
%Physik der Justus--Liebig--Universit\"at,
%D--35392 Giessen, Germany
}

\date{\today}
%\maketitle

\begin{abstract}
%The influence of an external magnetic field on the transport properties
%of an two-dimensional open quantum system is
%treated beyond the Markovian approximation.

The quantum   Langevin   formalism  is used
to  study  the  charge carrier transport in a two-dimensional sample. The
 center
of mass of   charge carriers is visualized as a quantum particle, while an environment
 acts as a heat bath coupled to it through the particle-phonon interaction.
%Solving the second order Heisenberg equations, the generalized Langevin equations are   obtained
%for quantum particle.
The dynamics of the charge carriers is limited
by the average collision time which takes effectively into account the two-body effects.
 The   functional dependencies   of particle-phonon interaction and   average collision
time  on the temperature and  magnetic field
are   phenomenologically treated.
 The galvano-magnetic
and thermo-magnetic effects  in the quantum system appear
  as the result of   the transitional processes
at low   temperatures.

\end{abstract}


\maketitle
\section{Introduction}

The behavior of solid matter under the influence of external
fields at   low temperature
is one of the interesting topics in solid-state physics \cite{Von,Kittel,Kittel2}. The
external field may be an electric field,   magnetic field,
optical signal, or temperature gradient. These   fields
modify some electronic properties, such as the carrier
concentration and the carrier mobility. Besides the carrier
mobility, the electric current is also affected by the  magnetic
field  which deflects its direction and leads to a nonzero cross voltage
(the classical Hall effect) linearly
proportional to the  field  strength \cite{Von,Kittel,Kittel2,Borovik}.
The oscillator
nature of the longitudinal magneto-resistance of bismuth sample at   low
temperature, known as the Shubnikov-De Haas effect, has been also observed   in the presence of
very intense magnetic fields \cite{Von,Kittel2,Komatsubara,Kah}.
The effect is more pronounced at low temperatures where the amplitude of oscillations is significantly
larger. The
first experimental study of the influence of electric fields of
the order of 100 mV cm$^{-1}$ on the Shubnikov-de Haas
magneto-resistance oscillations in $n$-InSb sample has been reported in Ref. \cite{Komatsubara}.
In addition to the shift of the extremes
to higher magnetic fields, a decrease of  oscillation
amplitudes with  electric field has been observed.
% However, the deviation from the linear law of Hall for
% the cross resistance has not been measured. The next revolutionary achievement
% in this field was the discovery of
 The Integer Quantum Hall Effect
(IQHE)  in the   GaAs-Al$_{0.3}$Ga$_{0.7}$As  heterostructure   has been
discovered~\cite{Klitz,Ebert} at strong external electromagnetic
fields and very low temperature.
%had an important impact on physics.
The quantization of
%conductance
conductivity   surprisingly occurred
in a certain two-dimensional electron gas (2DEG) under the influence of a
strong magnetic field.
In the
IQHE, the Hall conductance $\sigma_{xy}$  has a
stepwise dependence (the appearance of plateau) on the strong
external magnetic field. At these plateau, $\sigma_{xy}$ is
quantized as $\sigma_{xy}=i e^2 /(2 \pi \hbar)$, $i = 1, 2, . . .$,
% to an accuracy as high as several parts in$10^8$,
while the longitudinal conductivity $\sigma_{xx}$ nearly
vanishes. The vanishing   $\sigma_{xx}$ implies the absence of
dissipation.
%Note that the value of the
%transverse component $\rho_{xy}$ at the position of the plateaus
%of the steps is quantized to a few parts per billion to
%$\rho_{xy}=h/(ie^2)$, where "$i$" is an integer and $h$ is Planck's
%constant ($\rho_{xy}\approx 25.812 . . .  k\Omega$ for $i=1$).
%The value of the quantum of resistance, $h/e^2$,   as measured reproducibly
%to eight significant digits via the IQHE, became the world's new resistance standard.
%Concomitant with the quantization of $\rho_{xy}$ , the magneto-resistance
%$\rho_{xx}$ drops to vanishingly small values.
This is another
hallmark of the IQHE.
Also, the Fractional Quantum Hall Effect (FQHE),
where the Hall conductivity is quantized in fractional multiples of
$e^2 /(2 \pi \hbar)$, has been discovered~\cite{Tsui,Will}.

The theoretical models in Refs.  \cite{Pran,Laugh,Hald,Halp,Jain}
for describing the IQHE  and FQHE   have been developed.
The
combination of  a random potential  created by  impurities in a
sample  and  strong magnetic field gives rise to the special
coexistence of localized and extended electron states. As
known,   the Fermi level lies in the  energy gap (mobility gap) free
from the extended states and the change of the
electron density or the magnetic field can only result in
different occupations of localized states  which do  not affect
the conductivity. Based on these findings  the appearance of the conductivity
quantization has been explained.
%The explanation for the actual values
%of the quantized Hall conductance was given by an ingenious
%gedanken experiment, proving that a system with impurities has
%either zero conductance or the same conductance as a system with
%the same amount of electrons but without impurities.
The
general approach, which explains the quantization as well as the
integer quantized values, has been developed later with the
scaling theory. This approach also describes  correctly  the regions
where the conductivity is not quantized \cite{Pran}.
For the
theoretical explanation of the FQHE, the wave
functions have been introduced~\cite{Laugh} to describe the incompressible quantum states
and explain  the small but experimentally prominent class of
fractions $1/$odd. It turned out that the quasiparticle
excitations are  the charge/fluxcomposites with a fractional
charge and statistics, also known as  dubbed anions. The special properties of
charge/fluxcomposites have been used  in Refs.~\cite{Hald,Halp,Jain} to
construct  two so-called hierarchies, sets of Hall fractions
for which the incompressible ground states could be found. These
hierarchies are able to reproduce all  fractions  observed, but also
yield many fractions that have never been measured. The striking
universality in the manifestation of the quantum Hall effect
attracted large attention, not only
in solid state physics but also in high energy physics. The
extensive mathematical methods of topological field
theory~\cite{Bel} and infinite dimensional
algebras~\cite{Froh,Kac,Kar} have been applied to the IQHE and FQHE. As found in
several independent works, the description of
incompressible quantum states exploits the theory of chiral edge
currents~\cite{Froh}. The Quantum Hall ground states and
quasiparticle excitations have been described in terms of
representations of the infinite-dimensional
algebras~\cite{Froh,Kac,Kar}.

%Modeling the
%electric current implies the determination of the time-dependence of the number of electrons with the given momentum at certain
%location. The equations of motion for it can be obtained by using the quantum
%Langevin approach or density matrix formalism
%which is widely applied to find the effects of fluctuations and
%dissipation in macroscopic systems  \cite{Kampen1,Kampen2,Kampen3,Kampen4,Kampen5,Kampen6,LEG,DM,Katia,Isar}. The Langevin method in the kinetic theory
%significantly simplifies the calculation of nonequilibrium
%quantum and thermal fluctuations and provides a clear picture
%of the dynamics of the process \cite{Katia,Hu1,Hu2,Ford,Ford1,In,M110,Kanokov,Lac9,Lac10,Lac11,Lac12,Lac13,Lac14,Lac15,Lac16,Lac17}.

%The study of the solid matter at a very low temperature still
%remains a major issue in physics. Because of the influence of the
%very low temperature the properties of the system dramatically
%change and hence do not obey the conventional theories introduced
%at early stages.

%In given paper, we will focus mainly on
%the galvanomagnetic and thermomagnetic properties of solids.

%However there is still no general theory considering all type of
%observed phenomena together.

The aim of the present work is to
treat   the classic and quantum Hall effect as well as the Shubnikov-De Haas
effect  within the same model.
%in the external constant magnetic and electric fields
%beyond the Markov approximation (instantaneous dissipation and delta-correlated fluctuations)
%and the weak-coupling limit.
%In given paper, we   focus mainly on
%the galvanomagnetic and thermomagnetic properties of solids.
The basic idea of our model is the following. In the
electric current, we determine
the time-dependent   number of electrons with  given momentum at a certain
location.
We consider the center of mass of
  charge carriers with a positive charge $e=|e|$ as a quantum particle
  coupled to the environment (heat bath) through the
particle-phonon interactions. Solving  the second order
Heisenberg equations for the heat bath degrees of freedom, the
generalized non-Markovian Langevin equations are  explicitly
obtained for a quantum particle.
%At the same time,
The memory effects in these equations results from the coupling to
the environment.
%on the transport properties of the quantum particle, is also
%obtained without  any approximations for the particle-phonon  interaction.
The dynamics of the charge carriers is restricted
by the average collision time.
 The   functional form of the particle-phonon
coupling strength and the average collision time on the temperature and  magnetic field
are  phenomenologically   treated.

The paper is organized as follows.
In Sec.~II, we introduce the Hamiltonian of
the system and solve the
generalized non-Markovian Langevin equations for  a quantum particle.
 The electric and thermal conductivities  are derived in two-dimensional systems.
Note that the quantum
Langevin approach or the density matrix formalism
has been widely applied to find the effects of fluctuations and
dissipation in macroscopic systems
\cite{Kampen1,Kampen2,Kampen3,Kampen4,Kampen5,Kampen6,LEG,DM,Katia,Isar,Hu1,Hu2,Ford,Ford1,In,M110,Kanokov,Lac9,Lac10,Lac11,Lac12,Lac13,Lac14,Lac15,Lac16,Lac17,Lac18,Lac19}.
%The Langevin method in the kinetic theory
%significantly simplifies the calculation of nonequilibrium
%quantum and thermal fluctuations and provides a clear picture
%of the dynamics of the process \cite{Katia,Hu1,Hu2,Ford,Ford1,In,M110,Kanokov,Lac9,Lac10,Lac11,Lac12,Lac13,Lac14,Lac15,Lac16,Lac17}.
%The classical Hall effect in magnetic field  is considered
%, which appears as a result of a magnetic field.
The main assumptions of the model are discussed.
The  model   developed  is used in Sec. III   to  describe   the experimental
data on the classic and quantum Hall   and Shubnikov-De Haas
effects.  A summary  is given in Sec. IV.

\section{ Non-Markovian Langevin equations with external magnetic and electric fields}
\subsection{Derivation of quantum Langevin equations}
Let us consider   two-dimensional  motion of  a quantum charge particle in the presence
of heat bath and
 external constant electric  ${\bf E}=(E_x,0,0)$ and magnetic fields ${\bf B}=(0,0,B)$.
 The total
Hamiltonian of this system is
\begin{eqnarray}
H=H_c+H_b+H_{cb}.
\label{equ1}
\end{eqnarray}
The Hamiltonian $H_c$   describes the collective subsystem (quantum particle) with
effective mass tensor and charge $e=|e|$  in electric and magnetic fields:
\begin{eqnarray}
H_c
= \frac{1}{2 m_x}[p_x-e A_x(x,y)]^2+\frac{1}{2 m_y}[p_y-e
A_y(x,y)]^2+e E_x x= \frac{\pi_x^2}{2m_x}+\frac{\pi_y^2}{2m_y}+eE_x x.
%\nonumber
\label{equ2}
\end{eqnarray}
Here,  $m_x$ and $m_y$ are the components of the effective mass
tensor, ${\bf R}=(x,y,0)$ and
${\bf p}=(p_x,p_y,0)$ are the coordinate and canonically conjugated momentum,
respectively, ${\bf A}=(-\frac{1}{2}y B,\frac{1}{2}x B,0)$ is
 the vector potential of the magnetic field,
and the electric
field  $E_x$ acts in   $x$ direction. For simplicity, in Eq.~(\ref{equ2})
we introduce  the notations
$$\pi_x=p_x+\frac{1}{2}m_x\omega_{cx}y,\hspace{.3in}
\pi_y=p_y-\frac{1}{2}m_y\omega_{cy}x$$ with  frequencies
 $\omega_{cx}=\frac{e B}{m_x }$ and $\omega_{cy}=\frac{e
B}{m_y }$.
The cyclotron frequency is $\omega_c=\sqrt{\omega_{cx} \omega_{cy}}=\frac{e
B}{\sqrt{m_x m_y}}$.
% and $[\pi_x,\pi_y]=-[\pi_y,\pi_x]=i\hbar\mu\omega_c$.

The second term in Eq.~(\ref{equ1}) represents the Hamiltonian of the phonon
heat bath
% of the phonons,
\begin{eqnarray}
H_b=\sum_{\nu}^{}\hbar\omega_\nu b_\nu^+b_\nu,
\label{equ3}
\end{eqnarray}
where $b_\nu^+$ and $b_\nu$ are the phonon creation and
annihilation operators of the heat bath. The coupling between the
heat bath and   collective subsystem is described by
\begin{eqnarray}
H_{cb}=\sum_{\nu}^{}V_\nu({\bf R})(b_\nu^+ +
b_\nu)+\sum_{\nu}^{}\frac{1}{\hbar\omega_\nu}V_\nu^2({\bf R}).
\label{equ4}
\end{eqnarray}
The first term in  Eq. (\ref{equ4}) corresponds to the energy exchange  between
the collective subsystem and heat bath. We introduce the
counterterm (second term) in $H_{cb}$  to
compensate the coupling-induced renormalization of the collective potential.
Naturally, it can   be always separated from $e E_x x$ in
Eq.~(\ref{equ2}).
In general case, $V_\nu({\bf R})$ depends on the strength of
magnetic field and an impact of $\bf B$ is entered into the dissipative kernels
and random forces.

The equations of two-dimensional motion are
\begin{eqnarray}
\dot{x}(t)&=&\frac{i}{\hbar}[H,x]=\frac{\pi_x(t)}{m_x},\hspace{.3in}
\dot {y}(t)=\frac{i}{\hbar}[H,y]=\frac{\pi_y(t)}{m_y},\nonumber\\
\dot\pi_{x}(t)&=&\frac{i}{\hbar}[H,\pi_x]=\pi_{y}(t)\omega_{cy}-e
E_{x} -\sum_{\nu}^{}V'_{\nu,x}({\bf R})(b_\nu^+ + b_\nu)-
2\sum_{\nu}^{}\frac{V_{\nu}({\bf R})
V'_{\nu,x}({\bf R})}{\hbar\omega_{\nu}},\nonumber\\
\dot \pi_{y}(t)&=&\frac{i}{\hbar}[H,\pi_y]=-\pi_{x}(t)\omega_{cx}
-\sum_{\nu}^{}V'_{\nu,y}({\bf R})(b_\nu^+ + b_\nu)-
2\sum_{\nu}^{}\frac{V_{\nu}({\bf R})
V'_{\nu,y}({\bf R})}{\hbar\omega_{\nu}},
\label{equ5}
\end{eqnarray}
and
\begin{eqnarray}
\dot b_\nu^+(t)&=&\frac{i}{\hbar}[H,b_\nu^+]=
i\omega_\nu b_\nu^+(t) + \frac{i}{\hbar}V_{\nu}({\bf R}), \nonumber\\
\dot b_\nu(t)&=&\frac{i}{\hbar}[H,b_\nu]= -i\omega_\nu b_\nu(t)
- \frac{i}{\hbar}V_{\nu}({\bf R}).
\label{equ6}
\end{eqnarray}
The solution of Eqs.~(\ref{equ6}) are
\begin{eqnarray}
b_\nu^+(t)+b_\nu (t)&=&f^{+}_\nu (t)+f_\nu (t) -
\frac{2V_{\nu}({\bf R})}{\hbar\omega_\nu}  + \frac{2}{\hbar\omega_\nu}\int\limits_{0}^{t}d\tau
\dot V_{\nu}({\bf R}(\tau))\cos(\omega_\nu[t-\tau]),\nonumber\\
b_\nu^+(t)-b_\nu (t)&=&f^{+}_\nu (t)-f_\nu (t) +\frac{2i}{\hbar\omega_\nu}\int\limits_{0}^{t}d\tau
\dot V_{\nu}({\bf R}(\tau))\sin(\omega_\nu[t-\tau]),
\label{equ7}
\end{eqnarray}
where
\begin{eqnarray}
f_\nu(t)=[b_\nu(0)+\frac{1}{\hbar\omega_\nu}V_{\nu}({\bf R}(0))]e^{-i\omega_\nu t}. \nonumber
\end{eqnarray}
Substituting (\ref{equ7})
into (\ref{equ5}) and  eliminating the bath variables from
the equations of motion for the collective subsystem, we obtain the
set of nonlinear integro-differential stochastic dissipative
equations
\begin{eqnarray}
\dot {x}(t)&=&\frac{\pi_x(t)}{m_x}, \hspace{.3in}
\dot {y}(t)=\frac{\pi_y(t)}{m_y}, \nonumber\\
\dot \pi_{x}(t)&=&\pi_{y}(t)\omega_{cy} -e E_{x}-
 \frac{1}{2}\int\limits_{0}^{t}d\tau \{K_{xx}(t,\tau),\dot{x}(\tau)\}_{+}
 -\frac{1}{2}\int\limits_{0}^{t}d\tau \{K_{xy}(t,\tau),\dot{x}(\tau)\}_{+}+ F_{x}(t),\nonumber\\
\dot \pi_{y}(t)&=&-\pi_{x}(t)\omega_{cx} -
 \frac{1}{2}\int\limits_{0}^{t}d\tau \{K_{yy}(t,\tau),\dot{y}(\tau)\}_{+}
 -\frac{1}{2}\int\limits_{0}^{t}d\tau \{K_{yx}(t,\tau),\dot{y}(\tau)\}_{+}+ F_{y}(t).
\label{equ8}
\end{eqnarray}
The dissipative kernels and random forces in (\ref{equ8}) are
\begin{eqnarray}
K_{xx}(t,\tau)&=& \sum_{\nu}^{}\frac{1}{\hbar\omega_\nu}\{V'_{\nu,x}({\bf R}(t)),V'_{\nu,x}({\bf R}(\tau))\}_{+}\cos(\omega_\nu [t-\tau]), \nonumber\\
K_{xy}(t,\tau)&=& \sum_{\nu}^{}\frac{1}{\hbar\omega_\nu}\{V'_{\nu,x}({\bf R}(t)),V'_{\nu,y}({\bf R}(\tau))\}_{+}\cos(\omega_\nu [t-\tau]), \nonumber\\
K_{yx}(t,\tau)&=& \sum_{\nu}^{}\frac{1}{\hbar\omega_\nu}\{V'_{\nu,y}({\bf R}(t)),V'_{\nu,x}({\bf R}(\tau))\}_{+}\cos(\omega_\nu [t-\tau]), \nonumber\\
K_{yy}(t,\tau)&=&
\sum_{\nu}^{}\frac{1}{\hbar\omega_\nu}\{V'_{\nu,y}({\bf R}(t)),V'_{\nu,y}({\bf R}(\tau))\}_{+}\cos(\omega_\nu
[t-\tau])
\label{equ9}
\end{eqnarray}
and
\begin{eqnarray}
F_{x}(t)&=&\sum_{\nu}{}F_{x}^{\nu}(t)=-\sum_{\nu}{}V'_{\nu,x}({\bf R}(t))[f_{\nu}^{+}(t)+f_{\nu}(t)],\nonumber\\
F_{y}(t)&=&\sum_{\nu}{}F_{y}^{\nu}(t)=-\sum_{\nu}{}V'_{\nu,y}({\bf R}(t))[f_{\nu}^{+}(t)+f_{\nu}(t)],
\label{equ10}
\end{eqnarray}
respectively.
Here, we use the  notations: $V'_{\nu,x}=\partial V_{\nu}/\partial x$,
$V'_{\nu,y}=\partial V_{\nu}/\partial y$, and $\{Z_1,Z_2\}_+=Z_1Z_2+Z_2Z_1$.
Following the usual procedure of statistical mechanics,
we identify the  operators $F_{x}^{\nu}$ and $F_{y}^{\nu}$ as fluctuations because of the
uncertainty of the initial conditions for the bath operators. To
specify the statistical properties of   fluctuations, we
consider an ensemble of initial states in which the fluctuations
have the Gaussian distribution with zero average value
\begin{eqnarray}
\ll F^\nu_x (t)\gg = \ll F^\nu_y (t)\gg = 0.
\label{equ11}
\end{eqnarray}
The symbol $\ll...\gg$ denotes the  average over the bath with the
Bose-Einstein statistics
\begin{eqnarray}
\ll f_{\nu}^{+}(t)f_{\nu'}^{+}(t')\gg &=&\ll f_{\nu}(t)f_{\nu'}(t')\gg=0, \nonumber\\
\ll f_{\nu}^{+}(t)f_{\nu'}(t')\gg &=&\delta_{\nu,\nu'}n_{\nu}e^{i\omega_\nu [t-t']},\nonumber\\
\ll f_{\nu}(t)f_{\nu'}^{+}(t')\gg
&=&\delta_{\nu,\nu'}(n_{\nu}+1)e^{-i\omega_\nu [t-t']},
\label{equ12}
\end{eqnarray}
where the occupation numbers $n_\nu=[\exp(\hbar\omega_\nu/T)-1]^{-1}$ for phonons depend  on temperature
$T$ given in   energy units.

Using the properties   (\ref{equ11}) and (\ref{equ12}) of random forces,
% and neglecting terms of $\textit{O}(\hbar)$,
we obtain the following
symmetrized correlation functions $\varphi_{kk'}^{\nu}(t,t')=\ll
F^{\nu}_k(t)F^{\nu}_{k'}(t')+F^{\nu}_{k'}(t')F^{\nu}_k(t)\gg$
($k, k'=x, y$):
\begin{eqnarray}
\varphi_{xx}^\nu(t,t')&=&[2n_{\nu}+1]\{V'_{\nu,x}({\bf R}(t)),V'_{\nu,x}({\bf R}(t'))\}_{+}\cos(\omega_\nu [t-t']),\nonumber\\
\varphi_{yy}^\nu(t,t')&=&\varphi_{xx}^\nu(t,t')|_{{x\to y}},\nonumber\\
\varphi_{xy}^\nu(t,t')&=&[2n_{\nu}+1]\{V'_{\nu,x}({\bf R}(t)),V'_{\nu,y}({\bf R}(t'))\}_{+}\cos(\omega_\nu [t-t']),\nonumber\\
\varphi_{yx}^\nu(t,t')&=&\varphi_{xy}^\nu(t,t')|_{{x\to y}}.
%\varphi_{yx}^\nu(t,t')&=&[2n_{\nu}+1]\{V'_{\nu,y}({\bf R}(t)),V'_{\nu,x}({\bf R}(t'))\}_{+}\cos(\omega_\nu
%[t-t']).
\label{equ13}
\end{eqnarray}
The quantum fluctuation-dissipation relations read
\begin{eqnarray}
\sum_{\nu}^{}\varphi_{xx}^\nu(t,t')\frac{\tanh[\frac{\hbar\omega_\nu}{2T}]}{\hbar\omega_\nu}=K_{xx}(t,t'), \nonumber\\
\sum_{\nu}^{}\varphi_{yy}^\nu(t,t')\frac{\tanh[\frac{\hbar\omega_\nu}{2T}]}{\hbar\omega_\nu}=K_{yy}(t,t'),\nonumber\\
\sum_{\nu}^{}\varphi_{xy}^\nu(t,t')\frac{\tanh[\frac{\hbar\omega_\nu}{2T}]}{\hbar\omega_\nu}=K_{xy}(t,t'),\nonumber\\
\sum_{\nu}^{}\varphi_{yx}^\nu(t,t')\frac{\tanh[\frac{\hbar\omega_\nu}{2T}]}{\hbar\omega_\nu}=K_{yx}(t,t').
\label{equ14}
\end{eqnarray}
The validity of the fluctuation-dissipation relations means that we have properly identified the
dissipative terms in the non-Markovian dynamical equations of motion.
The quantum fluctuation-dissipation relations differ from the
classical ones and are reduced to them in the limit of high temperature.

\subsection{Solution of Non-Markovian Langevin equations}
In order to solve the equations of motion (\ref{equ8}) for the collective
variables, we apply the Laplace transformation. It significantly
simplifies the solution of the problem. After the Laplace
transformation, the equations of motion read
\begin{eqnarray}
x(s)s=x(0)+\frac{\pi_x(s)}{m_x}&,&\hspace{.2in}
y(s)s=y(0)+\frac{\pi_y(s)}{m_y}, \nonumber\\
\pi_{x}(s)s+\frac{\pi_{x}(s)}{m_x}(K_{xx}(s)+K_{xy}(s))&=&\pi_{x}(0)+\omega_{cy} \pi_{y}(s)-\frac{1}{s}e E_{x}+ F_{x}(s),\nonumber\\
\pi_{y}(s)s+\frac{\pi_{y}(s)}{m_y}(K_{yy}(s)+K_{yx}(w))&=&\pi_{y}(0)-\omega_{cx}
\pi_{x}(s)+ F_{y}(s).
\label{equ15}
\end{eqnarray}
Here,  $K_{xx}(s)$, $K_{yy}(s)$, $K_{xy}(s)$,   $K_{yx}(s)$ and $F_{x}(s)$, $F_{y}(s)$ are the Laplace transforms of the
dissipative kernels and random forces, respectively.
To solve these   equations, one should find the roots of the
  determinant
\begin{eqnarray}
D=s(m_x m_y \omega_c^2+[K_{xx}(s)+K_{xy}(s)+m_x s][K_{yy}(s)+K_{yx}(s)+m_y s])=0.
\label{equ16}
\end{eqnarray}
The explicit solutions for the originals are
\begin{eqnarray}
x(t)&=&x(0)+ A_{1}(t) \pi_{x}(0)+A_{2}(t)\pi_{y}(0)-A_{3}(t)eE_{x}+I_x(t)+I'_x(t) ,\nonumber\\
y(t)&=&y(0)+ B_{1}(t) \pi_{y}(0)-B_{2}(t) \pi_{x}(0)+B_{3}(t)eE_{x}- I_y(t)+I'_y(t), \nonumber\\
\pi_x(t)&=& C_{1}(t) \pi_{x}(0)+C_{2}(t)\pi_{y}(0)-C_{3}(t)eE_{x}+I_{\pi_x}(t)+I'_{\pi_x}(t), \nonumber\\
\pi_y(t)&=&D_{1}(t)\pi_{y}(0)-D_{2}(t)\pi_{x}(0)+D_{3}(t)eE_{x}-I_{\pi_y}(t)+I'_{\pi_y}(t),
\label{equ17}
\end{eqnarray}
where
$$I_x(t)=\int_0^t A_{1}(\tau)F_{x}(t-\tau)d\tau, \hspace{0.3in}
I'_x(t)=\int_0^t A_{2}(\tau)F_{y}(t-\tau)d\tau,$$
$$I_y(t)=\int_0^t
B_{2}(\tau)F_{x}(t-\tau)d\tau, \hspace{0.3in}   I'_y(t)=\int_0^t
B_{1}(\tau)F_{y}(t-\tau)d\tau,$$
$$I_{\pi_x}(t)=\int_0^t
C_{1}(\tau)F_{x}(t-\tau)d\tau, \hspace{0.3in} I'_{\pi_x}(t)=\int_0^t
C_{2}(\tau)F_{y}(t-\tau)d\tau,$$
$$I_{\pi_y}(t)=\int_0^t
D_{2}(\tau)F_{x}(t-\tau)d\tau, \hspace{0.3in} I'_{\pi_y}(t)=\int_0^t
D_{1}(\tau)F_{y}(t-\tau)d\tau$$
with the following time-dependent
coefficients:
\begin{eqnarray}
A_{1}(t)&=&\hat{L}^{-1}\left[\frac{K_{yy}(s)+K_{yx}(s)+m_y s}{D}\right]=B_{1}(t)|_{x\leftrightarrow y}, \nonumber\\
A_{2}(t)&=&m_y \omega_{cy} \hat{L}^{-1}\left[\frac{1}{D}\right]=B_{2}(t)|_{x\leftrightarrow y},\nonumber\\
A_{3}(t)&=&\hat{L}^{-1}\left[\frac{K_{yy}(s)+K_{yx}(s)+m_y s}{s
D}\right],\hspace{0.3in}
B_{3}(t)=m_x \omega_{cx}\hat{L}^{-1}\left[\frac{1}{s D}\right],\nonumber\\
C_{1}(t)&=&m_x \hat{L}^{-1}\left[\frac{s(K_{yy}(s)+K_{yx}(s)+m_y s)}{D}\right]=D_{1}(t)|_{x\leftrightarrow y},\nonumber\\
C_{2}(t)&=&m_x m_y \omega_{cy} \hat{L}^{-1}\left[\frac{s}{D}\right]=D_{2}(t)|_{x\leftrightarrow y},\nonumber\\
C_{3}(t)&=&m_x \hat{L}^{-1}\left[\frac{K_{yy}(s)+K_{yx}(s)+m_y
s}{D}\right],\hspace{0.3in} D_{3}(t)=m_x m_y \omega_{cx}
\hat{L}^{-1}\left[\frac{1}{D}\right].
\label{equ18}
\end{eqnarray}
Here, $\hat{L}^{-1}$ denotes the inverse Laplace transformation.
The exact solutions of $x(t)$, $y(t)$, $\pi_x(t)$, and $\pi_y(t)$
in terms of roots $s_i$ are given by the residue theorem.

For the system with linear coupling in coordinate, the coupling term is written as
\begin{eqnarray}
H_{cb}=\sum_{\nu}(\alpha_{\nu}x + \beta_{\nu}y)(b_\nu^+ + b_\nu)+
\sum_{\nu}\frac{1}{\hbar\omega_{\nu}}(\alpha_{\nu}x + \beta_{\nu}y)^{2},
\label{equ23}
\end{eqnarray}
where $\alpha_{\nu}$ and $\beta_{\nu}$ are the real coupling
constants. Here, we again introduce the counter term which depends
on the coordinates of   collective system and is treated as
a part of the potential. The operators of random forces and
dissipative kernels in Eqs. (\ref{equ8})   are
$$F_{x}(t)=-\sum_{\nu}{}\alpha_{\nu}(f_{\nu}^{+}+f_{\nu}),\hspace{.1in}
F_{y}(t)=-\sum_{\nu}{}\beta_{\nu}(f_{\nu}^{+}+f_{\nu})$$
and
\begin{eqnarray}
K_{xx}(t-\tau)&=& \sum_{\nu}^{}\frac{2\alpha_{\nu}^{2}}{\hbar\omega_\nu}\cos(\omega_\nu [t-\tau]), \nonumber\\
K_{yy}(t-\tau)&=&
\sum_{\nu}^{}\frac{2\beta_{\nu}^{2}}{\hbar\omega_\nu}\cos(\omega_\nu
[t-\tau]),
\label{equ24}
\end{eqnarray}
respectively.
We assume that there are no correlations between $F_x^\nu$
and $F_y^\nu$, so that $K_{xy}=K_{yx}=0$.
If the coupling constants $\alpha_{\nu}$ and $\beta_\nu$ depend on  magnetic field, then
the dissipative kernels $K_{xx}$ and  $K_{yy}$
are the functions of $B$.

It is convenient to introduce the spectral density $D_{\omega}$
of the  heat bath excitations  to replace the sum
over different oscillators  by an integral over the
frequency: $\sum_{\nu}^{}...\to \int\limits_{0}^{\infty}d\omega
D_{\omega}...$. This replacement is accompanied by the following
replacements: $\alpha_\nu\to\alpha_{\omega}$, $\beta_\nu\to\beta_{\omega}$,     $\omega_\nu\to\omega$, and $n_\nu\to n_{\omega}$.
Let us consider the following spectral functions \cite{Katia}
\begin{eqnarray}
D_{\omega}\frac{|\alpha_{\omega}|^2}{\hbar\omega}=
\frac{\alpha^2}{\pi}\frac{\gamma^2}{\gamma^2+\omega^2},\hspace{.3in}
D_{\omega}\frac{|\beta_{\omega}|^2}{\hbar\omega}=
\frac{\beta^2}{\pi}\frac{\gamma^2}{\gamma^2+\omega^2},
\label{equ25}
\end{eqnarray}
where the memory time $\gamma^{-1}$ of the dissipation  is inverse
to the phonon bandwidth of the heat bath excitations which are
coupled to a quantum particle. This is  the Ohmic
dissipation with the Lorentian cutoff (Drude dissipation)
\cite{Kampen1,Kampen2,Kampen3,Kampen4,Kampen5,Kampen6,Katia,Kanokov}.
%The relaxation time of the heat
%bath should be much less than the period of the collective
%system, i.e. $\gamma \gg \omega$.

Using the spectral functions (\ref{equ25}), we obtain the dissipative kernels and their
Laplace   transforms in the convenient form
\begin{eqnarray}
K_{xx}(t)&=&m_x \lambda_x\gamma e^{-\gamma|t|},\hspace{.3in}
K_{yy}(t)=m_y \lambda_y\gamma e^{-\gamma|t|},\nonumber\\
K_{xx}(s)&=&\frac{m_x \lambda_x\gamma}{s+\gamma},\hspace{.6in}
K_{yy}(s)=\frac{m_y \lambda_y\gamma}{s+\gamma},
%\nonumber\\
%K_{xx}(\omega)&=&\frac{m_x\lambda_x\gamma}{\gamma-i
%\omega},\hspace{.6in}
%K_{yy}(\omega)=\frac{m_y\lambda_y\gamma}{\gamma-i \omega},
%\hspace{.3in}(\Im [\omega] >0),
\label{equ26}
\end{eqnarray}
where
$$\lambda_x=\hbar \alpha^2=\frac{1}{m_x}\int_0^\infty K_{xx}(t-\tau)d \tau, \hspace{.3in}
\lambda_y=\hbar \beta^2=\frac{1}{m_y}\int_0^\infty K_{yy}(t-\tau)d
\tau$$
 are the friction
coefficients in the Markovian limit.
So,  the solutions for the collective
variables (\ref{equ17}) include the following time-dependent
coefficients:
\begin{eqnarray}
A_{1}(t)&=&\dot{A}_{3}(t),\hspace{.3in}
A_{2}(t)=\dot{B}_{3}(t)|_{x\leftrightarrow y},\nonumber\\
A_{3}(t)&=&\frac{1}{m_x}(\frac{\lambda_y}{\lambda_x\lambda_y+\omega_{c}^2}t+\frac{\omega_{c}^2(\gamma-\lambda_y)-\lambda_y^2(\gamma-\lambda_x)}{\gamma(\lambda_x\lambda_y+\omega_{cx}\omega_{cy})^2}\nonumber\\
&+&\sum_{i=1}^4 \frac{b_{i}e^{s_it}(\gamma+s_{i})(\gamma\lambda_y+s_{i}(\gamma+s_{i}))}{s_{i}^{2}}),\nonumber\\
B_{1}(t)&=&\dot{A}_3(t)|_{x\leftrightarrow y},\hspace{.3in}
B_2(t)=\dot{B}_{3}(t),\nonumber\\
B_{3}(t)&=&\frac{\omega_{cx}}{m_y}\left(\frac{t}{\lambda_x\lambda_y+\omega_{cx}\omega_{cy}}+\frac{2
\lambda_x\lambda_y-\gamma(\lambda_x+\lambda_y)}{\gamma(\lambda_x\lambda_y+\omega_{cx}\omega_{cy})^2}
+\sum_{i=1}^4 \frac{b_{i}e^{s_it}(\gamma+s_{i})^{2}}{s_{i}^{2}}\right),\nonumber\\
C_{1}(t)&=&m_x\ddot{A}_{3}(t),\hspace{.3in} C_{2}(t)=m_x
\ddot{B}_{3}(t),\hspace{.3in}
C_3(t)=m_x\dot{A}_3(t),\nonumber\\
D_1(t)&=&C_1(t)|_{x\leftrightarrow y},\hspace{.3in} D_2(t)=m_y
\ddot{B}_{3}(t),\hspace{.3in} D_3(t)=m_y\dot{B}_{3}(t),
\label{equ27}
\end{eqnarray}
where $b_i=[\prod_{j\neq i}(s_i-s_j)]^{-1}$ with $i,j=1,2,3,4$ and
$s_i$ are the roots of the  equation
\begin{eqnarray}
\gamma\lambda_x
[\gamma\lambda_y+s(\gamma+s)]+(\gamma+s)(s[s^{2}+\omega_{c}^2]
+\gamma[\omega_{c}^2+s(\lambda_y+s)])=0.
\label{equ28}
\end{eqnarray}

\subsection{Galvano-magnetic effects}

In order to determine the transport coefficients, we use Eqs.
(\ref{equ17}). Averaging them over the whole system and by
differentiating in $t$, we obtain the  system of equations for
the first moments
\begin{eqnarray}
<\dot{x}(t)>&=&\frac{<\pi_x(t)>}{m_x},\hspace{.3in}
<\dot{y}(t)>=\frac{<\pi_y(t)>}{m_y},\nonumber\\
<\dot{\pi}_x(t)>&=&\tilde{\omega}_{cy}(t)<\pi_y(t)>-\lambda_{\pi_x}(t)<\pi_x(t)>-e \tilde{E}_{xx}(t),\nonumber\\
<\dot{\pi}_y(t)>&=&-\tilde{\omega}_{cx}(t)<\pi_x(t)>-\lambda_{\pi_y}(t)<\pi_y(t)>-e\tilde{E}_{xy}(t),
\label{equ29}
\end{eqnarray}
with the friction coefficients
\begin{eqnarray}
\lambda_{\pi_{x}}(t)=-\frac{D_{1}(t)\dot{C}_{1}(t)+D_{2}(t)\dot{C}_{2}(t)}{C_{1}(t)D_{1}(t)+C_{2}(t)D_{2}(t)},
\nonumber\\
\lambda_{\pi_{y}}(t)=-\frac{C_{1}(t)\dot{D}_{1}(t)+C_{2}(t)\dot{D}_{2}(t)}{C_{1}(t)D_{1}(t)+C_{2}(t)D_{2}(t)},
\label{equ30}
\end{eqnarray}
and   renormalized cyclotron frequencies
\begin{eqnarray}
\tilde{\omega}_{cx}(t)=\frac{D_{1}(t)\dot{D}_{2}(t)-D_{2}(t)\dot{D}_{1}(t)}{C_{1}(t)D_{1}(t)+C_{2}(t)D_{2}(t)},
\nonumber\\
\tilde{\omega}_{cy}(t)=\frac{C_{1}(t)\dot{C}_{2}(t)-C_{2}(t)\dot{C}_{1}(t)}{C_{1}(t)D_{1}(t)+C_{2}(t)D_{2}(t)},
\label{equ31}
\end{eqnarray}
while the components of the electric field read:
\begin{eqnarray}
\tilde{E}_{xx}(t)=E_{x}[D_{3}(t)\tilde{\omega}_{cy}(t)+C_{3}(t)\lambda_{\pi_{x}}(t)+\dot{C}_{3}(t)],\nonumber\\
\tilde{E}_{xy}(t)=E_{x}[C_{3}(t)\tilde{\omega}_{cx}(t)-D_{3}(t)\lambda_{\pi_{y}}(t)-\dot{D}_{3}(t)].
\label{equ32}
\end{eqnarray}
As seen,
the dynamics is governed by the non-stationary coefficients.  As found,   the external
magnetic field generates the flow of charge carriers and electric field  in the cross
direction (the classical  Hall effect).
It should be noted that the cross component $\tilde{E}_{xy}(t)$
of the electric field is  initially absent
and appears during the non-Markovian evolution of the collective subsystem.

Let us consider the magneto-transport process in the two-dimensional
 system with current density
  defined as \cite{Von,Kittel,Kittel2}
\begin{eqnarray}
J_{i}=\sum_{j=1}^{2}\sigma_{ij}({\bf B})E_j.
\label{equ33}
\end{eqnarray}
Here, $\sigma_{ij}({\bf B})$ is the electric conductivity tensor which
depends on the magnitude and direction of the magnetic field
${\bf B}$. One   can also define the current density by
using the expression for the collective momentum (\ref{equ17}) and the
fact that ${\bf J}=-ne\dot{{\bf R}}$ because
\begin{eqnarray}
J_{x}=\frac{ne^2}{m_j}C_3(t)E_x(t),\hspace{.3in}
J_{y}=-\frac{ne^2}{m_y}D_3(t)E_x(t).
\label{equ34}
\end{eqnarray}
If we change the direction of the external electric field
$\vec{E}(E_x,0,0)$ to $\vec{E}(0,E_y,0)$, the expression for the
current density (\ref{equ34}) change to
\begin{eqnarray}
J_{x}=\frac{ne^2}{m_x}\tilde{D}_3(t)E_y(t),\hspace{.4in}
J_{y}=\frac{ne^2}{m_y}\tilde{C}_3(t)E_y(t),
\label{equ35}
\end{eqnarray}
where
%the coefficients $\tilde{D}_3(t)$ and $\tilde{C}_3(t)$ are
%obtained Eqs. (\ref{equ8}) with
%switched external electric field $E_y$:
$$\tilde{C}_{3}(t)=m_y L^{-1}\left[\frac{K_{xx}(s)+K_{xy}(s)+m_x s}{D}\right],\hspace{.9in}
\tilde{D}_{3}(t)=m_x m_y \omega_{cy}
L^{-1}\left[\frac{1}{D}\right].$$

Comparing (\ref{equ33}) with (\ref{equ34}) and (\ref{equ35}),
one can write the expression for   conductivity tensor at time $t=\tau$
\begin{eqnarray}
\sigma(\tau)=n e^2
  \begin{pmatrix}
    \frac{C_3(\tau)}{m_x} &&& -\frac{D_3(\tau)}{m_y} \\
    \frac{\tilde{D}_3(\tau)}{m_x} &&& \frac{\tilde{C}_3(\tau)}{m_y}
  \end{pmatrix},
\label{condten}
\end{eqnarray}
while its   inverse transformation yields the specific resistance tensor
\begin{eqnarray}
\rho(\tau)= \frac{1}{n
e^2[C_3(\tau)\tilde{C}_3(\tau)+D_3(\tau)\tilde{D}_3(\tau)]} \begin{pmatrix}
  m_x \tilde{C}_3(\tau) &&& m_x D_3(\tau) \\
 -m_y \tilde{D}_3(\tau) &&& m_y C_3(\tau)
 \end{pmatrix}.
\label{resistten}
\end{eqnarray}
The non-diagonal elements of the specific magneto-resistance tensor
have the meaning of the Hall resistance
\begin{eqnarray}
\rho_H(\tau)=\frac{m_x D_3(\tau)}{n
e^2[C_3(\tau)\tilde{C}_3(\tau)+D_3(\tau)\tilde{D}_3(\tau)]}= \frac{m_y
\tilde{D}_3(\tau)}{n e^2[C_3(\tau)\tilde{C}_3(\tau)+D_3(\tau)\tilde{D}_3(\tau)]}.
\label{resisttenroh}
\end{eqnarray}
In the case of two charge carriers, the model  is generalized in Appendix A.

\subsection{The main assumptions of the model}
Here, we list the main assumptions of the model which allow us to proceed with the
calculations for real systems.
We suppose that in each collision  the charge carriers lose their ordered motion and
 their velocities  vanish. As in  the kinetic theory of gases, in our model
  we assume that the lengths and times $t=\tau$ of free path
  are the same for all charge carriers and all collisions.
So, we introduce the time limit  $t=\tau$ in the conductivity
tensor (\ref{condten}) or   resistance tensor (\ref{resistten}).
In our model there are three different
characteristic times describing the dynamics of  charge carriers:
1) the relaxation time $\tau_r={\lambda}^{-1}$ ($\lambda=\lambda_x=\lambda_y$),
2) the average time between two collisions $\tau$, and
3) the memory time ${\gamma}^{-1}$ of the heat bath excitations.
The values of $\tau_r$  and $\tau$ are   related with   one-body (mean-field)
and two-body effects (dissipations). So, by introducing  the time parameter $\tau$, we
take effectively into consideration
the two-body collisions of charge carriers. The mean free
time $\tau$ is related to the thermodynamic equilibrium properties of
the material, whereas the relaxation time $\tau_r$ relates to the thermal
and electrical transport properties (see Fig.~1).
The relaxation
time  $\tau_r$ of electrons  is the characteristic time for a
distribution of charge carriers in a solid to approach or "relax" to
equilibrium after the disturbance is removed. A familiar example is
the relaxation of
current   to   zero
equilibrium value after the external  electric field  is turned off.
%The
%conductivity of a material is directly proportional to this
%relaxation time;
Highly conductive materials have relatively long
relaxation and free motion times.
%The average distance that an electron travels
%before a collision is called the mean free path.
At $\tau\gg\tau_r$, the one-body (mean-field)  dissipation dominates.
If these times are comparable, then the process
has a transitional behavior.
Note that in general the values of
$\tau_r$ and $\tau$ depend on  temperature $T$ and the strength of magnetic field $B$.
\begin{figure}[h!]
  \includegraphics[width=10cm]{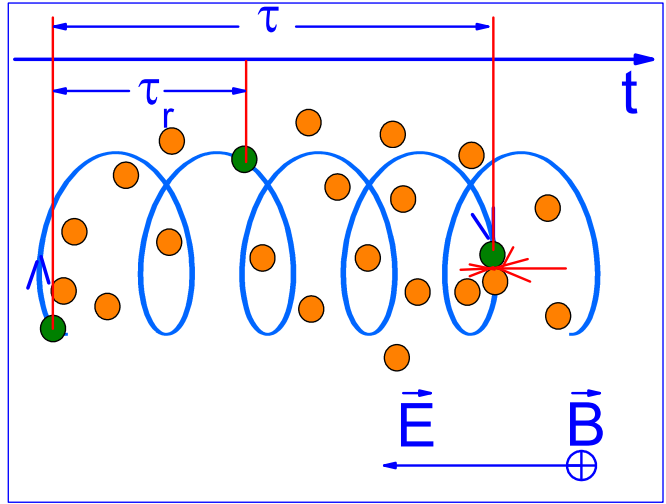}\\
  \caption{Schematic presentation of the scattering process and different time scales in a two-dimensional magneto-transport. }
\label{1_fig}
\end{figure}

\subsection{Axial symmetric system}
One can obtain  clearer physical picture of the process,
if the space-symmetric system is considered. In this system
$m_x=m_y=m$, $\lambda_x=\lambda_y=\lambda$, and
$\omega_{cx}=\omega_{cy}=\omega_c$. So, Eqs. (\ref{equ28}), which defines the poles,
%in the integrands of $I_j$ and $I'_j$ ($j=x,y,\pi_x,\pi_y$),
is simplified:
\begin{eqnarray}
(s^2+\omega_c^2)(\gamma+s)^2+2\gamma \lambda s
(\gamma+s)+\lambda^2 \gamma^2=0.
\label{equ33}
\end{eqnarray}
This equation has the
roots
\begin{eqnarray}
s_1&=&-\frac{1}{2}\left(\gamma+i \omega_c+\sqrt{(\gamma-i
\omega_c)^2-4 \gamma \lambda}\right),\hspace{.3in}
s_2=s_1^*,\nonumber\\
s_3&=&-\frac{1}{2}\left(\gamma+i \omega_c-\sqrt{(\gamma-i
\omega_c)^2-4 \gamma \lambda}\right),\hspace{.3in}
s_4=s_3^*.\nonumber
%\label{equ33}
\end{eqnarray}
In order to split the real and imaginary parts of the roots, we
expand them up to the first order in $\lambda/\gamma$:
\begin{eqnarray}
s_1&=&-\frac{\lambda
\gamma^2}{\gamma^2+\omega_c^2}-i\frac{\omega_c^2+\gamma^2+\lambda
\gamma}{\gamma^2+\omega_c^2}\omega_c, \nonumber\\
%s_2&=&-\frac{\lambda \gamma^2}{\gamma^2+\omega_c^2}+i\frac{\omega_c^2+\gamma^2+
%\lambda \gamma}{\gamma^2+\omega_c^2}\omega_c,\nonumber\\
s_3&=&-\gamma\frac{\gamma^2+\omega_c^2-\gamma
\lambda}{\gamma^2+\omega_c^2}+i\frac{\lambda \gamma
\omega_c}{\gamma^2+\omega_c^2}.\nonumber
%s_4&=&-\gamma\frac{\gamma^2+\omega_c^2-\gamma
%\lambda}{\gamma^2+\omega_c^2}-i\frac{\lambda \gamma
%\omega_c}{\gamma^2+\omega_c^2}.\nonumber
%\label{equ34}
\end{eqnarray}
%
%In this approximation, according to (2.1.27) components of
%conductivity equal:
%\begin{eqnarray}
%\sigma_{xx}(t)&=&\frac{n e^2}{m(\lambda^2+\omega_c^2)}\bigg(\lambda\\
%&+&\left.\exp[-\frac{\lambda \gamma^2 t}{\gamma^2+\omega_c^2}]
%\left(\omega_c \sin[\frac{(\gamma^2+\gamma\lambda+\omega_c^2)\omega_c t}{\gamma^2+\omega_c^2}]-
%\lambda \cos[\frac{(\gamma^2+\gamma\lambda+\omega_c^2)\omega_c t}{\gamma^2+\omega_c^2}]\right)\right),\nonumber\\
%\sigma_{xy}(t)&=&\frac{n e^2}{m(\lambda^2+\omega_c^2)}\bigg(\omega_c\nonumber\\
%&-&\left.\exp[-\frac{\lambda \gamma^2 t}{\gamma^2+\omega_c^2}]
%\left(\lambda
%\sin[\frac{(\gamma^2+\gamma\lambda+\omega_c^2)\omega_c
%t}{\gamma^2+\omega_c^2}]+\omega_c
%\cos[\frac{(\gamma^2+\gamma\lambda+\omega_c^2)\omega_c
%t}{\gamma^2+\omega_c^2}]\right)\right).\nonumber
%\end{eqnarray}
%
%
%\section{Comparison of our model with Drude model}
%
In this approximation, the components of
conductivity (\ref{condten}) at  $t=\tau$ are
\begin{eqnarray}
\sigma_{xx}(\tau)&=&\frac{\sigma_{xx0}}{1+(\omega_c \tau_r)^2}\nonumber\\
&-&\frac{\sigma_{xx0}}{\sqrt{1+(\omega_c
\tau_r)^2}}\exp[-\frac{\gamma^2
}{\gamma^2+\omega_c^2}\frac{\tau}{\tau_r}]
\cos[\frac{(\gamma^2+\gamma/\tau_r+\omega_c^2)\omega_c \tau}{\gamma^2+
\omega_c^2}+\arctan(\frac{1}{\omega_c \tau_r})],\nonumber\\
\sigma_{xy}(\tau)&=&\frac{\sigma_{xx0}\omega_c \tau_r}{1+(\omega_c \tau_r)^2}\nonumber\\
&-&\frac{\sigma_{xx0}}{\sqrt{1+(\omega_c
\tau_r)^2}}\exp[-\frac{\gamma^2
}{\gamma^2+\omega_c^2}\frac{\tau}{\tau_r}]
\cos[\frac{(\gamma^2+\gamma/\tau_r+\omega_c^2)\omega_c
\tau}{\gamma^2+\omega_c^2}-\arctan(\omega_c \tau_r)].
\label{equasym}
\end{eqnarray}
%where we introduced
%the relaxation  time $\tau_r={\lambda}^{-1}$, and the average time  $\tau$ between two collisions.
As seen, the expressions for
macroscopically observable values such as the cross and longitudinal
components of conductivity (resistance) contain the non-oscillatory and
oscillatory parts. In very strong magnetic fields, when the cyclotron frequency is much
larger than the friction of the system $\omega_c\gg \tau_r^{-1}$,
the longitudinal and transverse components of conductivity
oscillate in antiphase.
 Depending on the  ratio between $\tau_r$ and $\tau$, the oscillatory or
non-oscillatory term of (\ref{equasym}) has a major role.
At $\tau\gg \tau_r$ or $\tau\to\infty$, the oscillatory term vanishes and we obtain  the
Drude conductivity
\begin{eqnarray}
\sigma=\frac{\sigma_{xx0}}{1+(\omega_c \tau_r)^2}
  \begin{pmatrix}
    1 &&& -\omega_c \tau_r \\
    \omega_c \tau_r &&&
    1
  \end{pmatrix},
\end{eqnarray}
where
$$\sigma_{xx0}=\frac{n e^2 \tau_r}{m} $$
is the
Drude conductivity
at $B=0$.
As seen, the conductivity   (\ref{equasym})  differs from the
Drude one at $B\ne 0$ by additional oscillatory term.

%%% ----------------------------------------------------------------------

\subsection{Thermomagnetic effects}
%
%In the previous subsections we
%considered the galvanomagnetic
%processes and worked out the method
%for finding the conductivity and resistance tensors.
%
Here, we   assume that the electric and thermal
currents are carried by the same particles
%called free electrons in the conventional theory
and find the relation between the
electric and thermal conductivities.
%Contrary to the electric
%current, i.e. to the electric charge transfer, not only the free
%electrons participate in heat transfer but also phonons as well.
%So, two parts of the thermal conductivity,
%electron and phonon conductivities, are separated. Though it is supposed that the
%electron conductivity much dominates the phonon conductivity with
%the exception of the case of very low temperatures when the metal
%contains many kind of dopands.
If the electric energy gradient
$eE$ in the   Hamiltonian (2) is substituted by the
temperature gradient $\frac{dT}{dx}$ (the temperature is in
energy units), the generating force of   particle motion is
changed from the electric to  thermal potential, giving rise to  the
 thermomagnetic effects. Taking
into consideration the expression for the heat flux
$${\bf Q}=n \varepsilon_{kin}\dot{{\bf R}},$$
%where $\vec{v}=\dot{{\bf R}}$ is the velocity of the charge carrier,
and
 following  the procedure of subsection II.C, we   find the expressions for the components
of thermal conductivity tensor
\begin{eqnarray}
\chi(\tau)
%=n  \varepsilon_{kin}(\tau)
%  \begin{pmatrix}
%    \frac{C_3(\tau)}{m_x} &&& -\frac{D_3(\tau)}{m_y}\\
%    \frac{\tilde{D}_3(\tau)}{m_x} &&& \frac{\tilde{C}_3(\tau)}{m_y}
%  \end{pmatrix}
  =\frac{1}{e^2}\varepsilon_{kin}(\tau)\sigma(\tau),
\end{eqnarray}
where  $\varepsilon_{kin}$ is the kinetic energy of charge
carriers.
%The time dependencies of
The kinetic energy is defined through
the variances $\Sigma_{\pi_x\pi_x}$ and $\Sigma_{\pi_y\pi_y}$ (Appendix B) and   mean values $<\pi_x>$ and
$<\pi_y>$.
%
%\begin{eqnarray}
%\varepsilon_{kin}(\tau)=\frac{<\pi_x^2(\tau)>}{2m_x}+\frac{<\pi_y^2(\tau)>}{2m_y}=
%\frac{\Sigma_{\pi_x\pi_x}(\tau)+<\pi_x(\tau)>^2}{2m_x}+\frac{\Sigma_{\pi_y\pi_y}(\tau)+\pi_y(\tau)>^2}{2m_y}.
%\end{eqnarray}
%The terms of
%$<\pi_x(t)>$ and
%$<\pi_y(t)>$ depend
%on the initial Gaussian distribution. These terms can always be
%set zero by choosing appropriate initial conditions. So, the
%collective energy mainly depend on variances
%$\Sigma_{\pi_x\pi_x}(t)$ and $\Sigma_{\pi_y\pi_y}(t)$ (Appendix D).
%
In the
quasi-equilibrium high temperature limit ($\tau\rightarrow\infty$), the kinetic energy
\begin{eqnarray}
\varepsilon_{kin}(\infty)=T
\end{eqnarray}
is
defined by  the equipartition theorem (the equilibrium variances $\Sigma_{\pi_x\pi_x}(\infty)=m_x T$ and
$\Sigma_{\pi_y\pi_y}(\infty)=m_y T$).
Finally, one can  rederive the classical Wiedemann-Franz law \cite{Von,Kittel,Kittel2}
\begin{eqnarray}
L=\frac{\chi}{T\sigma}=\frac{1}{e^2}=const,\nonumber
\end{eqnarray}
where $L$ is  the Lorentz number
which reflects the fact that the
ability of the
%quasiparticles
carriers to carry a charge is the same as to
transport heat.

% Now it can be read through the electrical
%conductivity tensor and variances as
%\begin{eqnarray}
%\chi(t)=\frac{\sigma(t)}{e^2}
%\left(\frac{\Sigma_{\pi_x\pi_x}(t)}{2m_x}+\frac{\Sigma_{\pi_y\pi_y}(t)}{2m_y}\right)
%  .\nonumber
%\end{eqnarray}

\section{Calculated results and discussions}
\begin{figure}[h!]
    \includegraphics[width=12cm]{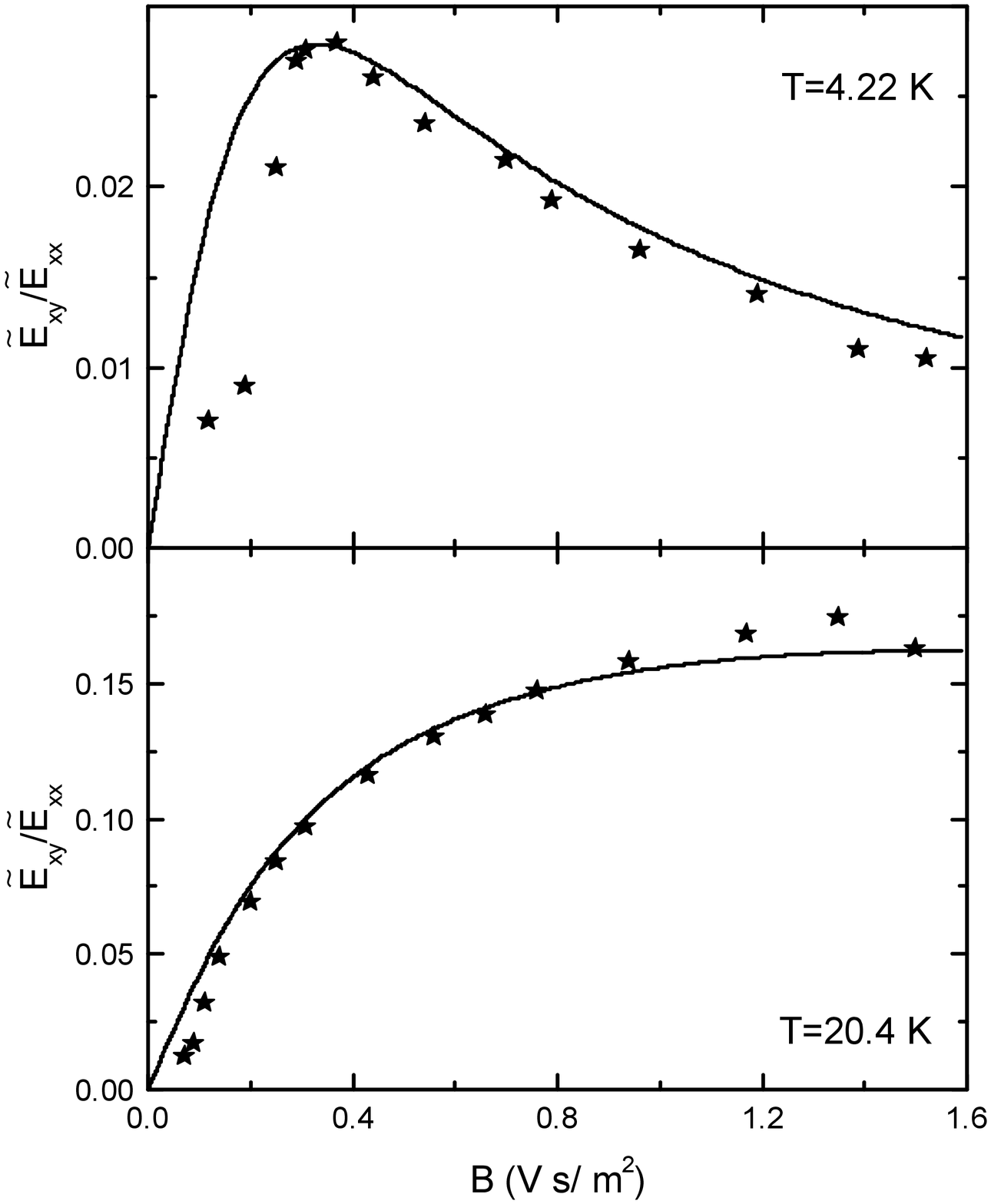}
\caption{The experimental~\cite{Borovik} (symbols) and theoretical dependencies (lines) of the
    tangent of the Hall angle on magnetic field  $B$
    for zinc at   temperatures $T$ indicated.}
\label{2_fig}
\end{figure}
In the calculations we set $\lambda=\lambda_x=\lambda_y$ (or $\lambda_\pi=\lambda_{{\pi}_x}=\lambda_{{\pi}_y}$)
and
$m=m_x=m_y$ (or $\omega_{c}=\omega_{cx}=\omega_{cy}$).
In order to turn to the observable values, all parameters
$\tau^{-1}$,   $\tau_r^{-1}$,  $\omega_c$, and $\gamma$ in
the  expressions are multiplied by     $\frac{m}{e}$:
\begin{eqnarray}
\tau^{-1} &\to& \frac{m}{e}\tau^{-1},\\
\tau_r^{-1} &\to& \frac{m}{e}\tau_r^{-1}=\mu^{-1},\nonumber\\
\omega_c &\to& \frac{m}{e}\omega_c=B,\nonumber\\
\gamma &\to& \frac{m}{e}\gamma=\Gamma.\nonumber
\end{eqnarray}
As a result, instead of the friction coefficient $\lambda$,
cyclotron frequency $\omega_c$, and inverse response time $\gamma$ of the system
 we  have the inverse reciprocal mobility $\mu^{-1}$ of charge carriers,
 intensity of the magnetic field $B$, and new parameter
$\Gamma$ connected to the memory time.
The   mobility of a charge carrier $\mu_0=\mu(B=0)$ in the
absence of   magnetic field  is the measurable value. The value of $B$ is set
by the experimental condition.
%The parameter $\Gamma$ is
%connected to the memory time.

In addition, one can  also
study the magnetic moment of the system.
It should be noted  that in
our model  the influence of the magnetic field on the
coupling between quantum particle and heat-bath is neglected.
The impact of
magnetic field is entered into the dissipative kernels.
%
%explain an experimentally detected increase in magnetoresistance
%of some elements \cite{Fred, Sond, Kohler}. The influence of the
%magnetic field is entered into the kernels
%(2.2.4).
However, there are  solids with constant
resistance     in the wide spectrum of magnetic
field. Their properties   can be described by
neglecting the effect of magnetic field on the coupling term.

\subsection{Classical Hall effect}
The classical case corresponds  $\tau\gg\tau_r$ or $\tau\to\infty$.
To demonstrate the capabilities of the model, we calculate  the
tangent of the Hall angle,
$\tan[\Theta_H]=\tilde{E}_{xy}(\infty)/\tilde{E}_{xx}(\infty)$, for the sample of
Zn settled in the increasing external magnetic field at two
temperatures. Many experiments were performed to measure this value
in several materials. We choose Zn \cite{Borovik}
because it has one type of charge carriers, and
consequently,  the technique of implementation of
the model can  be easily   understood.
%
%In the case of matters with more then one type of charge
%carriers the problem is more complicated, since the two-band model
%given above should be considered. In order to turn to the
% observable values in expressions (31), all parameters in
%these expressions should be multiplied by the  mass to charge ratio ($m/e$).
%As a result, instead of the friction coefficient $\lambda$,
%cyclotron frequency $\omega_c$, and inverse response time $\gamma$ of the system
%we  have the inverse reciprocal mobility of charge carriers
%1/$\mu$, intensity of the magnetic field $B$ and new parameter
%$\Gamma=m \gamma/e$.
%
%From the experimental data~\cite{Borovik} we
%may define the strength $B$ of the magnetic field at which the charge
%carriers deviate to the maximal angle from their non-field
%direction. Knowing the field, we define the parameter $\Gamma$
%by using expression (37).
%
The calculated and experimental
characteristics of Zn are listed in Table I. The calculations performed with the
values of mobility $\mu_0$   at $B=0$  are in a good agreement with the
experimental data (Fig.~2), especially at high strength of magnetic field.
\begin{table}[h!]
    \caption{Experimental (asterisks)~\cite{Borovik} and theoretical
    characteristics of Zn at two temperatures.
    The value of  $B_{max}^{\ast}$ corresponds to the position of the maximum of
      experimental non-diagonal component of electric field as a function of magnetic field.}\label{1}
    \begin{flushleft}
  \begin{tabular*}{\textwidth}{@{\extracolsep{\fill}}       c   c  c  c  c}
    \hline
%    \hline
     Temperature$^\ast$,& Resistance$^\ast$, $\rho_{xx},$& Mobility, & $\Gamma=B_{max}^{\ast},$ &Max. Hall  \\
      $T$  (K)& $\times 10^{-11}$ ($\Omega\cdot$ m) &$\mu_0$ (m$^2$ / V$\cdot$ s )  & (Tesla) & angle$^\ast$, $\Theta_H$    \\ \hline
      4.22& 2.6555 & 50.25 & 0.37 & $1.6\,^{\circ}$   \\
      20.4 & 35.595& 1.1 & 1.35 & $9.37\,^{\circ}$  \\
    \hline
%    \hline
   \end{tabular*}
  \end{flushleft}
\end{table}
\begin{figure}[t]
  \includegraphics[width=12cm]{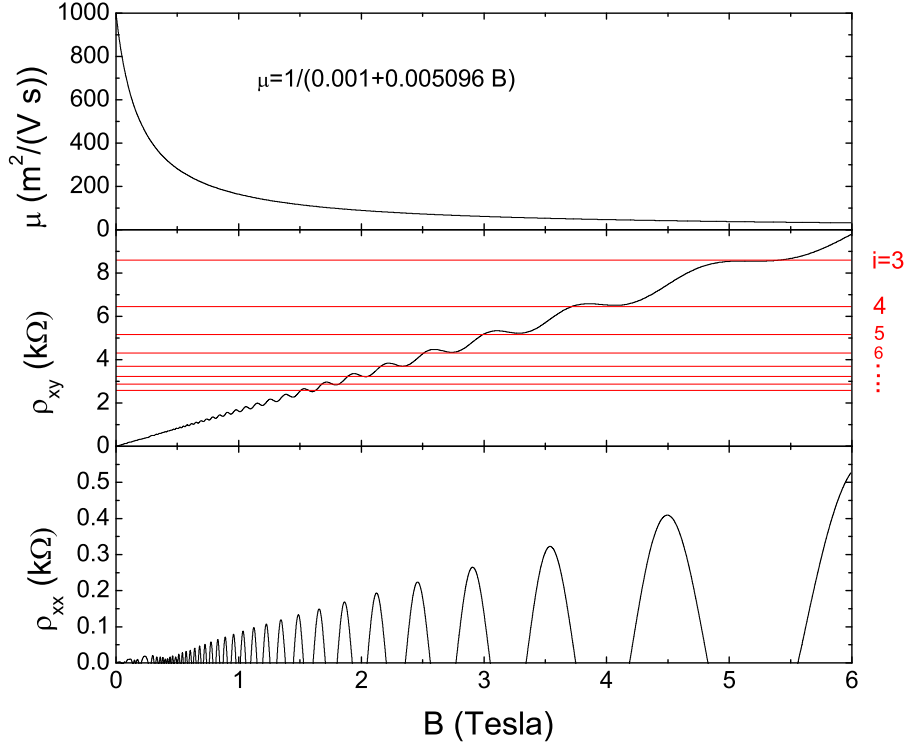}\\
  \caption{Theoretical plots of the components of resistivity tensor and
  mobility of charge carriers in  GaAs-Al$_{0.3}$Ga$_{0.7}$As  at $T=50$ mK. The experimental
  concentration, $n=3.7\cdot 10^{15}$ m$^{-2}$, is   from Ref. \cite{Ebert}
  and the mobility has the
  functional form $\mu(B)=1/(0.001+0.005096 B)$ indicated on the plot.
  The ratio of   mean collision time per relaxation time is set to be
  $\frac{\tau}{\tau_r}=2$, and $\Gamma=100/\mu_0$.}
\label{3_fig}
\end{figure}
\begin{figure}[h!]
      \includegraphics[width=12cm]{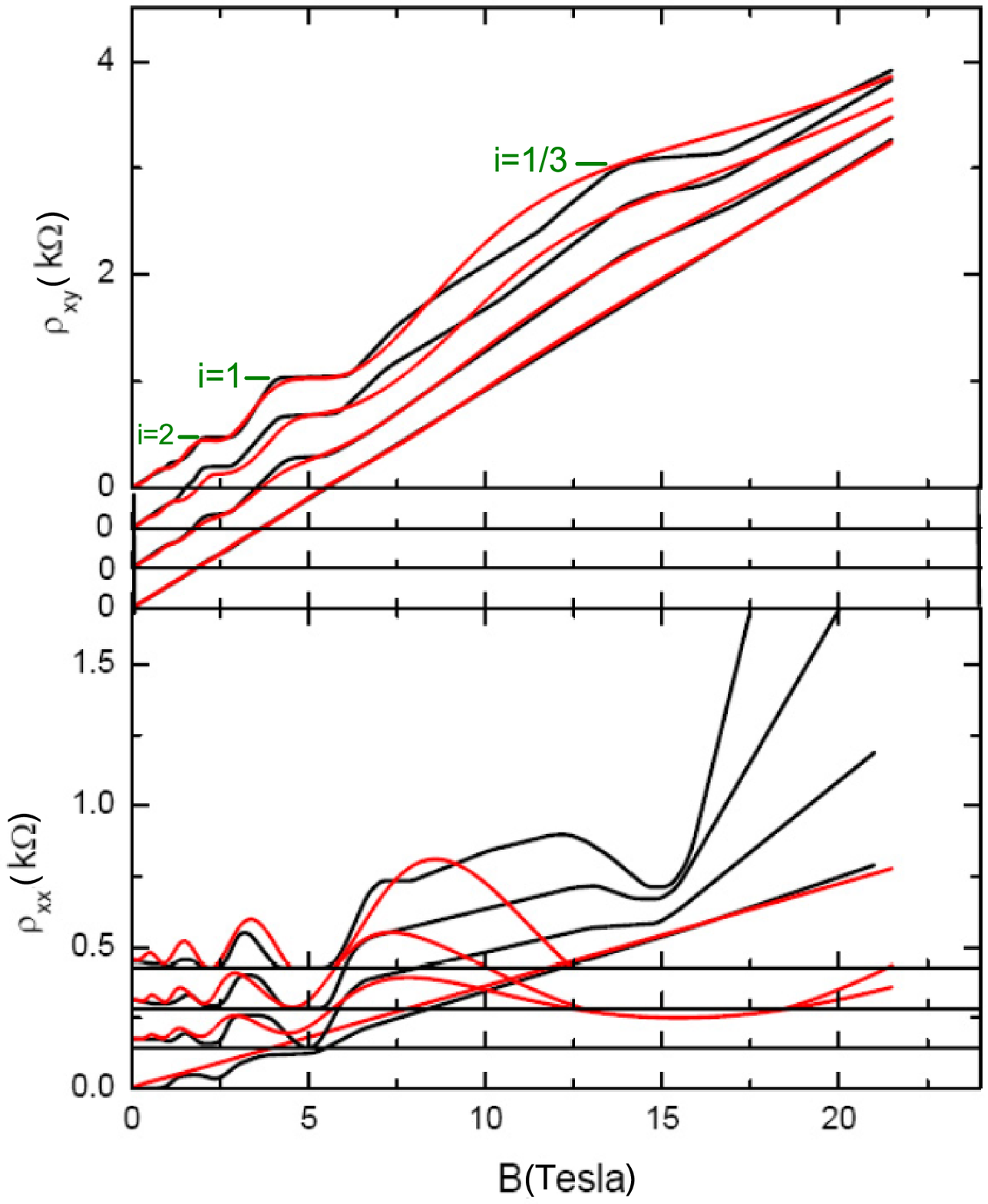}
%      \begin{minipage}[b]{7cm}
    \caption{The experimental (black)~\cite{Tsui} and theoretical (red) curves of magnetic field
  dependencies of the components of resistivity tensor at different temperatures $T$=0.48, 1.00, 1.65,
  4.15 K in the order from   top to
  bottom.
%  \vspace{5cm}
  }
%  \end{minipage}
\label{4_fig}
 \end{figure}

\subsection{Integer Hall effect}
%\subsubsection{Experimental situation}
%\subsubsection{Theoretical interpretation}
The IQHE has been observed
in the   heterostructure   GaAs-Al$_{0.3}$Ga$_{0.7}$As    at the low temperature of $T=50$ mK~\cite{Ebert}.
According to the experiment, the concentration of the
artificially prepared two-dimensional sample  GaAs-Al$_{0.3}$Ga$_{0.7}$As
is  $n=3.7\cdot 10^{15}$ m$^{-2}$. There are large intervals of $B$
where the longitudinal conductivity has its minimum,
while the Hall conductivity is quantized with immense precision in
integer multiples of $e^2/(2\pi\hbar)$.
The calculated components
of the resistivity tensor are shown in Fig. 3 for wide range of magnetic field.
The observed increase of the width of plateau in $\rho_{xy}$ with the field is
 explained by the decrease of the mean collision time of
charge carriers. The
external magnetic field effects   the coupling between the
collective system and heat-bath. This coupling  linearly rises
  with the magnetic field which induces the reciprocal decrease of
  $\tau_r$. Such  a decrease in relaxation time
effectively changes the mobility of charge carriers (Fig.~3).
The drastic decrease of  mobility can be attributed to
the effect of localization under the influence of magnetic field.
For the formation of the step-wise nature of $\rho_{xy}$, the
ratio $\tau/\tau_r$ is required to be constant at   whole
magnetic field spectrum. Thus, both the mean collision time  and
relaxation time  fall down inversely with increasing magnetic field, as in the experiment  \cite{Von}. However,
 the ratio between them remains
constant. As seen in  Fig. 3,
 in the region between two plateaus for $\rho_{xy}$ the longitudinal resistivity
has the maximum, while at the center of  plateau it is
minimal. This phenomenon is explained by the $\pi/2$ phase
shift in the oscillations of $\rho_{xx}$ and $\rho_{xy}$ which is
clearly visible in the approximate  formulas
for the axial symmetric system at very strong magnetic fields ($\omega_c\tau_r\gg 1$):
\begin{eqnarray}
\frac{\rho_{xx}}{\rho_{xx0}}&=&1-\mu B\exp\left[-\frac{\Gamma^2
}{\Gamma^2+B^2}\frac{\tau}{\tau_r}\right]
\cos\left[\frac{(\Gamma^2+\Gamma/\mu+B^2)(e/m) \tau B}{\Gamma^2+B^2}\right],\nonumber\\
\frac{\rho_{xy}}{\rho_{xx0}}&=& \mu B\left(1-\exp\left[-\frac{\Gamma^2
}{\Gamma^2+B^2}\frac{\tau}{\tau_r}\right]
\sin\left[\frac{(\Gamma^2+\Gamma/\mu+B^2)(e/m) \tau
B}{\Gamma^2+B^2}\right]\right),\nonumber
\end{eqnarray}
where $\rho_{xx0}=1/\sigma_{xx0}$.
%
%%% ----------------------------------------------------------------------
%\goodbreak

\subsection{Fractional Hall effect}
\begin{figure}[h!]
  \includegraphics[width=12cm]{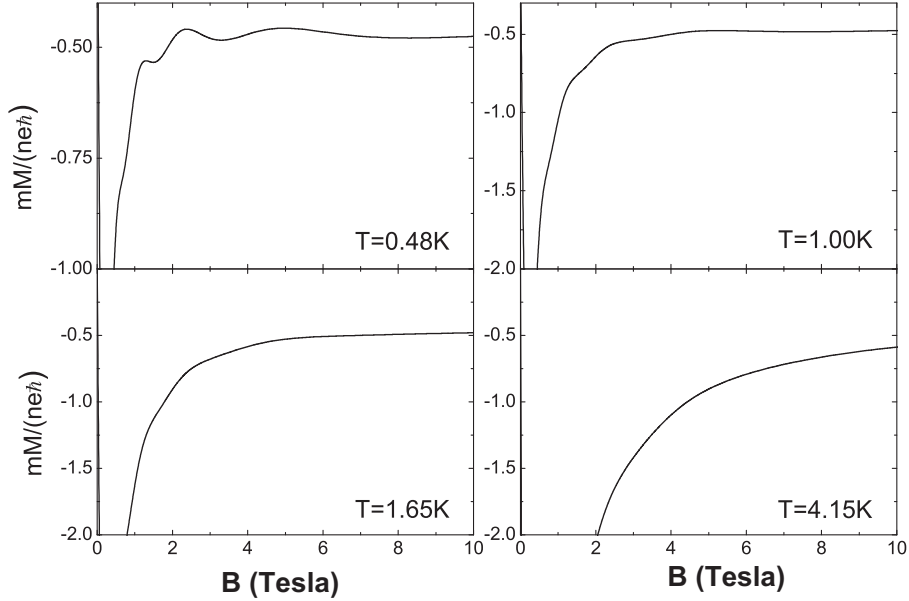}
  \caption{The calculated magnetic moment as a function of magnetic field at indicated temperatures.}
\label{5_fig}
\end{figure}
\begin{figure}[h!]
      \includegraphics[width=12cm]{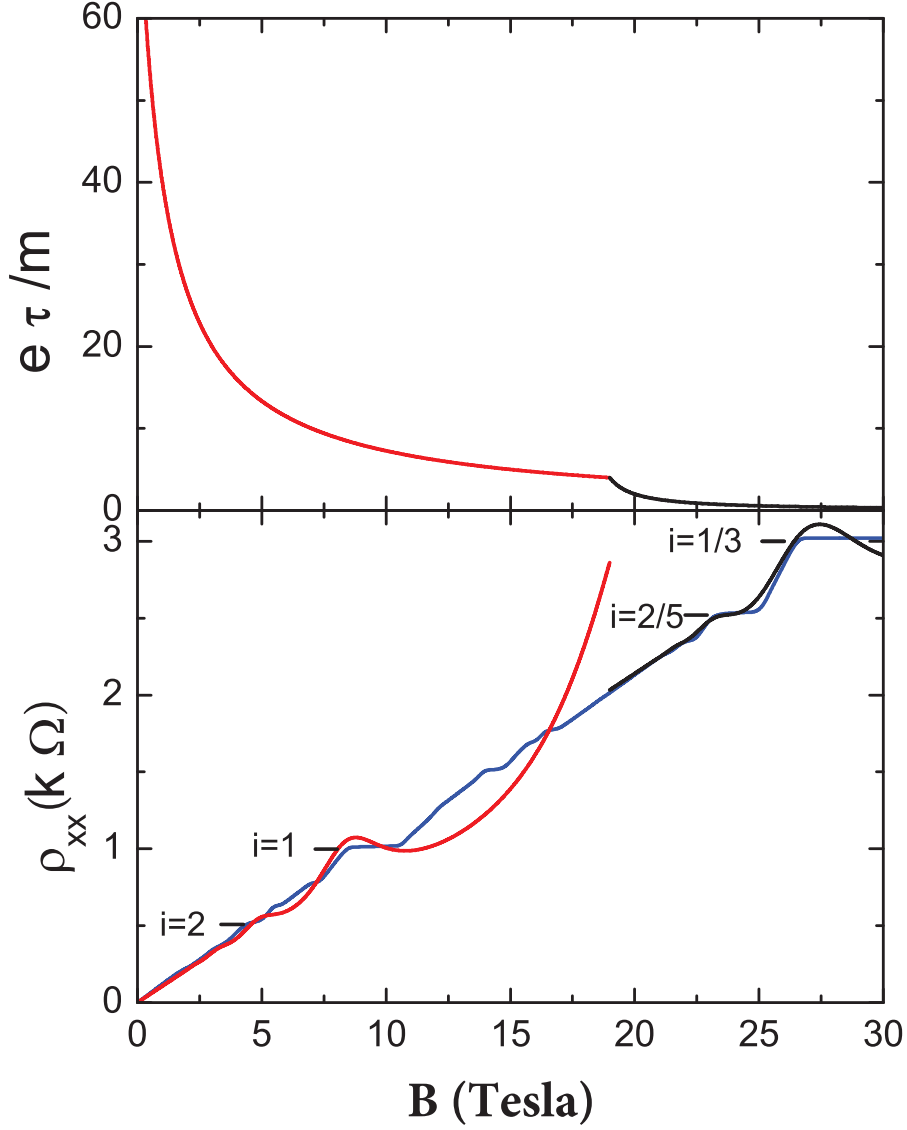}
%      \begin{minipage}[b]{5cm}
    \caption{The calculated dependence of transverse component of resistivity and the mean free time on magnetic
    field. The blue line corresponds to the experiment~\cite{Will}.
%    \vspace{5cm}
}
%  \end{minipage}
\label{6_fig}
 \end{figure}
 %
%
%\subsubsection{Experiment and theoretical interpretation}
%The discovery of "anomalous" quantum Hall effect took the
%condensed matter community completely by surprise.
The  Hall plateau  at strong magnetic field has been discovered in  Ref.~\cite{Tsui}  (Fig. 4) and
corresponds to the fractional value of the filling factor $i=1/3$.
The experiment has been carried out at four temperatures below the
helium temperature for the sample of  GaAs-Al$_{0.3}$Ga$_{0.7}$As  with
the 2D concentration $n=1.23\cdot 10^{15}$ m$^{-2}$ and carrier
mobility $\mu_0=9$ m$^2$/(V $\cdot$ s). The step-wise appearance of
the Hall resistivity becomes smoother with increasing temperature.
The purity of the sample is so high that the electrons move
ballistically, i.e. without scattering against impurity atoms,
over relatively long distances.
\begin{table}[h!]
    \caption{The experimental (asterisks)~\cite{Tsui} and theoretical parameters used in the calculations of the FQHE.
    MF denotes a magnetic field.
    }
    \label{1}
    \begin{flushleft}
  \begin{tabular*}{\textwidth}{@{\extracolsep{\fill}}  c    c   c  c  c  c}
    \hline \hline
    Temperature$^*$,& Mobility in the absence& Ratio & Functional form of  \\
    $T$  (K)& of MF, $\mu_0^*$  (m$^2$ / V$\cdot$ s)  & $\tau/\tau_r$  &  $\mu(B)=\tau_r(B)e/m$  \\ \hline
    0.48& 9 & 1.95 & $(0.11+0.056 B)^{-1}$  \\
    1.00& 9 & 2.84 & $(0.11+0.064 B)^{-1}$  \\
    1.65& 9 & 3 & $(0.11+0.084 B)^{-1}$  \\
    4.15& 9 & 4.5 & $(0.11+0.185 B)^{-1}$  \\
    \hline \hline
   \end{tabular*}
  \end{flushleft}
\end{table}
In our  calculations (Fig. 4), we take the experimental values of
mobility $\mu_0$ and 2D concentration $n$ and
$\Gamma=100/\mu_0$ (Table II). As in the case of the integer Hall effect,   the
relaxation time and mean collision time of charge carriers decrease
inversely with increasing magnetic field and their ratio remains constant (Table II).

We
also calculate  the magnetic moment $M(\tau)=\frac{n e L_z(\tau)}{2m}$, where
\begin{eqnarray}
L_z(\tau)&=&<x(\tau)\pi_y(\tau) - y(\tau)\pi_x(\tau)>\nonumber\\
&=&\frac{m\hbar\gamma^2}{\pi}
\int_0^{\infty}\int_0^{\tau}\int_0^{\tau} \frac{d\omega dt dt'\omega \coth\left[\frac{\hbar\omega}{2T}\right]}{\omega^2+\gamma^2}\cos(\omega[t-t']) \nonumber\\
&\times& \left\{\lambda_x[B_2(t)C_1(t')-A_1(t)D_2(t')]+\lambda_y[A_2(t)D_1(t')-B_1(t)C_2(t')]\right\}
\label{390}
\end{eqnarray}
is the $z$-component of  angular momentum.
% (here $m_x=m_y=m$ or $\omega_{cx}\omega_{cy}=\omega_c$ and $\lambda_x=\lambda_y=\lambda$).
The calculations were performed with the
parameters from Table II. As seen in Fig.~5, the magnetic moment approaches to
$\frac{n e}{2m}\hbar$ at all   temperatures considered.
At   low
temperatures $T$=0.48 and 1 K,   small oscillations of the
magnetic moment are observed in the region of weak magnetic field.
These oscillations are getting smoother as the temperature increases and
disappear at sufficiently high temperatures.
%Note that at
% $\omega_c\gg \sqrt{\lambda_x \lambda_y}$, $\gamma\to\infty$, and $T\to 0$,
% the angular momentum and magnetic moment are quantized:
%\begin{eqnarray}
%\label{42}
%L_z=-\hbar,  \hspace{0.2 in}    M=-\frac{n e \hbar}{2m}.\nonumber
%\end{eqnarray}

In Fig.~6, one can see the  experimental curve (blue line)
obtained for the  sample GaAs/AlGaAs  at lower  temperature
85 mK. The measured concentration of the 2DEG created in this sample
is $n=3\times10^{15}$ m$^{-2}$ and the mobility of charge carriers at $B=0$
is $\mu_0=100$ m$^2$/V$\cdot$s. According to this figure the width of the
plateau  increases up to 10 Tesla and then   suddenly decreases.
It starts to increase again  from
19 Tesla to 30 Tesla. It is obvious
that above  10 Tesla the properties
of the system  drastically change. In  our model this
behavior   is explained by the abrupt change of the
functional form of the mean collision time. Initially being
$$\frac{e}{m}\tau=\frac{80}{1.3+B}$$
below 19 Tesla (red line),   it changes
to $$\frac{e}{m}\tau=\frac{4}{B-18}$$ above 19 Tesla (black line). The parameter
$\Gamma$  is  again  set to be $100/\mu_0$. It should be
noted that   the mobility $\mu=e\tau_r/m=\mu_0$ at
$T=85$ mK remains constant
in a whole range of magnetic field.

%%% ----------------------------------------------------------------------

As shown,  the non-oscillatory
term of the conductivity
  (resistance) plays a key role at  high temperature (the classical Hall effect), whereas an
oscillatory part   mainly contributes to the resistance at   low
temperature (the quantum Hall and Shubnikov-De Haas effects), where the mobility of charge carriers is sufficiently
large and the values of   relaxation time and average collision time are comparable. One should stress
that  the Shubnikov-De Haas, integer and fractional quantum Hall effects are the results of the transitional processes.
Note that  for the integer and fractional quantum Hall effects,  the values of
relaxation time and average collision time should be comparable.

\subsection{Shubnikov-De Haas effect}
\begin{figure}[h!]
  \includegraphics[width=16cm]{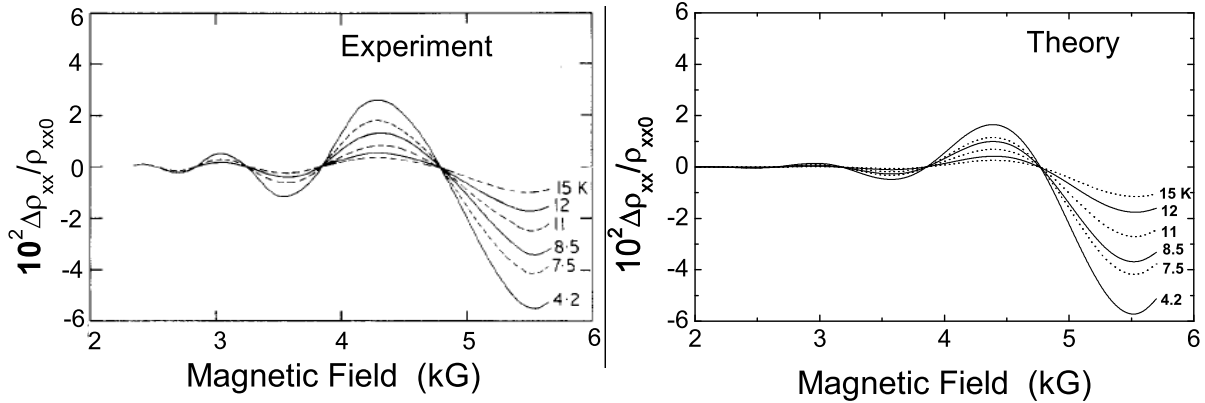}\\
  \caption{The experimental~\cite{Kah} and theoretical dependencies of
  the oscillatory part of the longitudinal magneto-resistance on
magnetic field for  various temperatures.}
\label{7_fig}
\end{figure}
\begin{figure}[h!]
      \includegraphics[width=12cm]{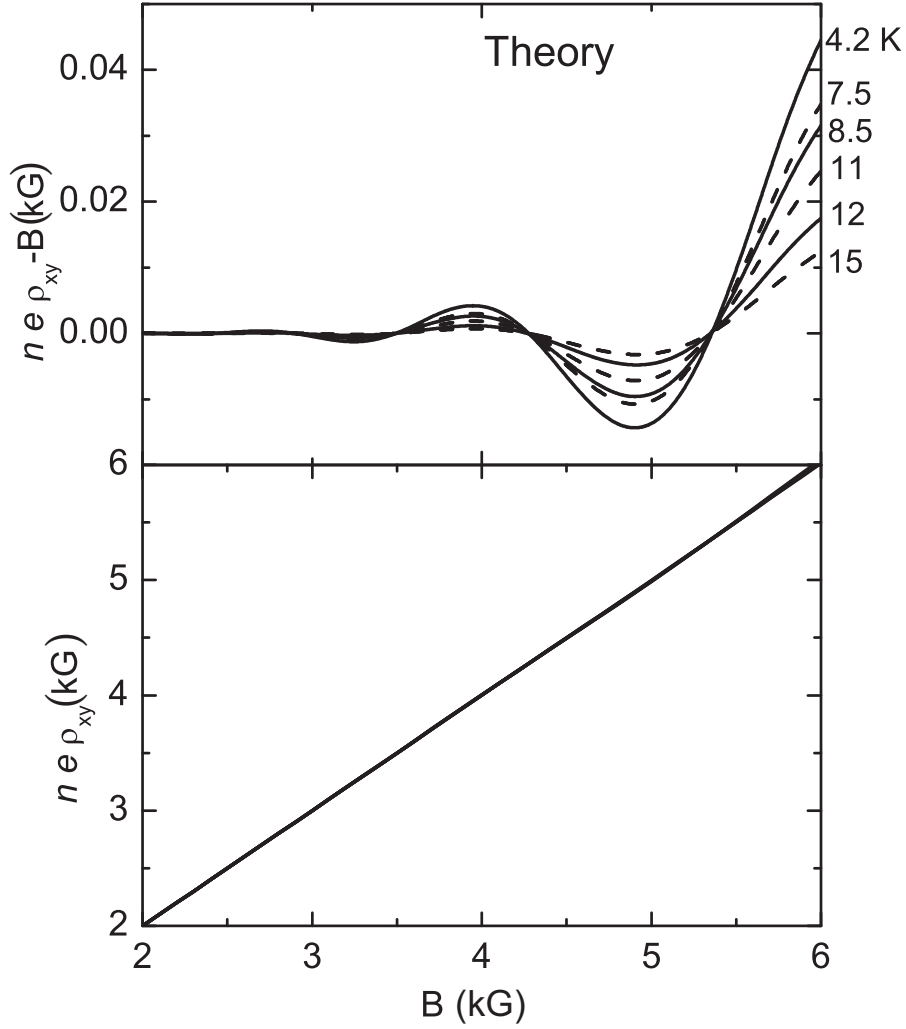}
%      \begin{minipage}[b]{7cm}
    \caption{The dependence  of the calculated cross resistance on
magnetic field at   various temperatures.
%\vspace{6cm}
}
%  \end{minipage}
\label{8_fig}
 \end{figure}
\begin{figure}[h!]
      \includegraphics[width=12cm]{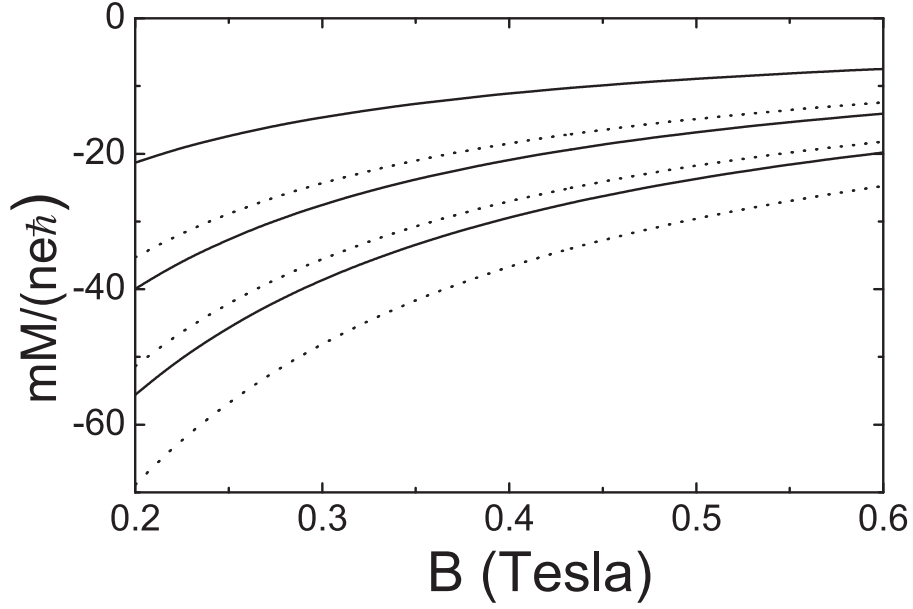}
%      \begin{minipage}[b]{7cm}
    \caption{The calculated magnetic field dependence of the magnetic moment. The curves from   top to
  bottom correspond to the calculations at temperatures $T$=4.2, 7.5, 8.5, 11, 12, and 15 K.
%  \vspace{2cm}
}
%  \end{minipage}
\label{9_fig}
 \end{figure}
\begin{figure}[h!]
  \includegraphics[width=12cm]{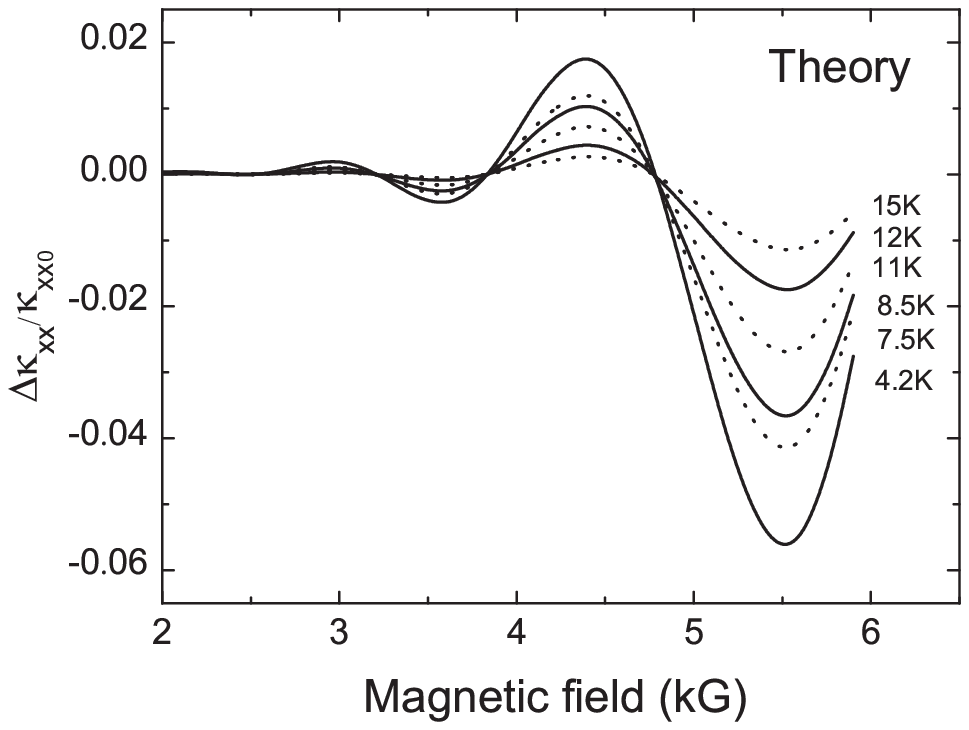}\\
  \caption{The calculated dependencies of
  the oscillatory part of the longitudinal thermal-resistance on
magnetic field for  various temperatures.}
\label{10_fig}
\end{figure}
%
%\subsubsection{Experimental facts and theoretical interpretation}
%The oscillations of the conductivity of a material (Shubnikov-De
%Haas effect) that occurs at low temperatures in the presence of
%very intense magnetic fields \cite{Kittel2}. The amplitude of oscillations is
%larger at low temperatures than that at high temperatures. The
%first experiment concerning the influence of electric fields of
%the order of 100 $mV cm^{-1}$ on the Shubnikov-de Haas (ShD)
%magneto-resistance oscillations in $n-InSb$ was performed in Ref. \cite{Komatsubara}.
%In addition to a shift of the extrema
%to higher magnetic fields, a decrease of the oscillatory
%amplitudes with  electric field was observed.

Let us consider the experiment performed in Ref.~\cite{Kah} with the   $n$-InSb  sample which
has been prepared from the $n$-type single crystal of InSb having a charge
carrier concentration   $n=5.9 \cdot 10^{15}$ cm$^{-3}$ (see Fig.~7).
%\subsection{Theoretical interpretation}
According to the experiment, the mobility of charge
carriers in the absence of   magnetic field varies within 5\%
with increasing temperature  from 4.2 to 15 K. However, this value
does not show a variation with   increasing magnetic field.
%It remains constant in the whole range of considered magnetic field at  fixed temperature.
The points of intersection between the zero
axis and the resistance curves     are the same
  at all   temperatures.
This implies that the period of oscillations does not  depend on
temperature. Moreover, the equal increase in period of
oscillations with the magnetic field is observed at all
temperatures in the experiment.
%Based on the theoretical calculations, one can conclude
%that
So, contrary to the
cases of integer and fractional Hall effects, the mobility $\mu$ or
relaxation time $\tau_r$ of charge carriers in the   sample   does not
change with   magnetic field  while the  mean
free time decreases similarly  as the
magnetic field increases at any temperature. Thus, the ratio
of the mean collision time and the relaxation  time, which remains
constant in the quantum Hall regimes, now  decreases with increasing
field (see Table III and Fig.~7).
% ------------------------------------------------------------------------
\begin{table}[h!]
    \caption{The experimental (asterisks)~\cite{Kah} and
    theoretical parameters used in the calculations of Shubnikov-De Haas effect.
    MF denotes a magnetic field.
    }\label{1}
    \begin{flushleft}
  \begin{tabular*}{\textwidth}{@{\extracolsep{\fill}}  |c|c|c|c|c|c|c|}
    \hline \hline
    Temperature$^*$,  $T$  (K)& 4.2 & 7.5 & 8.5 & 11 & 12 & 15 \\ \hline
    Mobility in the absence  of MF, $\mu_0^*$  (m$^2$ / V$\cdot$ s)  & 9.5 & 9.1 & 8.8 & 8.4 & 8.3 & 8 \\ \hline
    Functional form of $\tau(B)e/m$ & \multicolumn{6}{c|} {1000/(1.9+12.8 B)}     \\ \hline \hline
    $\Gamma$ & \multicolumn{6}{c|} {100/$\mu_0$}     \\ \hline \hline
   \end{tabular*}
  \end{flushleft}
\end{table}
No oscillations have been detected in the magnetic field
dependence of transverse resistance in the experiment.
Though  there are    oscillations in our
approach  (Fig.~8),   their
amplitudes are negligible with respect to the non-oscillatory part
of the resistance. In order to detect them, the non-oscillatory
part should be subtracted from the measured value as $\Delta
\rho_{xy}=\rho_{xy}-\frac{B}{n e}$. The calculations of the
magnetic moment do not show the oscillations in Fig.~9.
The absolute value of   magnetic moment decreases with increasing magnetic
field.

In Fig. 10, the dependencies of the longitudinal thermal resistance
$k_{xx}=\frac{\chi_{xx}}{\chi^2_{xx}+\chi^2_{xy}}$ on magnetic
field are presented at different temperatures. Comparing Figs. 7 and 10,
one can notice the correlations between magneto-  and thermal-resistances.

\section{Summary}
Using the non-Markovian Langevin approach
and    coupling between the charge carriers and environment,
the behavior of the  generated flow of charge carriers under the influence of
external magnetic field was investigated for the   two-dimensional
case.
   The   model developed was applied to the case where the collective
    coordinates are linearly coupled   to the
heat-bath coordinates.  In order to average the influence of  environment  on the
collective system, we applied the spectral function
of   heat-bath excitations which describes the Drude
dissipation with Lorenzian cutoffs.
 The dynamics of  charge carriers was limited
by the average collision time as in the kinetic theory of gases. In this way,
the two-body effects were taken effectively into consideration.
 The  functional dependencies of the average collision time and
coupling strength between the charge carriers and
environment on   temperature and   magnetic field
were phenomenologically treated.
One can say that we solve the inverse problem by finding
suitable coupling strengths and the average collision times
for describing the experimental data.
As shown, the galvano- and thermo-magnetic effects
  strongly depend on the ratio between the relaxation time (the inverse friction coefficient)
and  average collision time.

The explicit   expressions were obtained for the macroscopically
observable values such as transverse and longitudinal components of conductivity
  (resistance)  and the Hall angle.
It was concluded that the non-oscillatory
term of conductivity
  (resistance) plays a key role at  high temperature (the classical Hall effect) whereas the
oscillatory part of conductivity
  (resistance) mainly contributes to the resistance at   low
temperature (quantum Hall and Shubnikov-De Haas effects), where the mobility of charge carriers is sufficiently
large and the values of   relaxation time and average collision time are comparable. Thus,
the Shubnikov-De Haas, integer and fractional quantum Hall
effects are the results of the transitional processes.

 The Shubnikov-De Haas effect has been observed
both in the two-dimensional and three-dimensional samples, whereas
the integer and fractional quantum Hall effects have been detected
only in the two-dimensional samples so far. However, our model
also predicts their existence  in the three-dimensional samples.
The model was applied to the thermomagnetic
processes as well. The oscillations of the thermal coefficients were
predicted in the quantum Hall and Shubnikov-De Haas regimes.
The experimental observation of such oscillations would be a good
criteria for the justification of the present  model.

The  model  developed can be extended further  by taking
the spin of electrons and non-stationary external
fields into consideration.

%According to the experimental reports the Shubnikov-De Haas effect
%has been observed both in the 2D and 3D samples, whereas the
%integer and fractional quantum Hall effects have been detected
%only in the 2D samples so far. However our model predicts their
%existence also in the 3D samples. The oscillations of thermal
%coefficients were also predicted at quantum Hall and Shubnikov-De
%Haas regimes.

\acknowledgments
This work was partially supported by  the Russian Foundation for Basic Research (Moscow)   and  DFG (Bonn).
The IN2P3(France)-JINR(Dubna) Cooperation
Programme is gratefully acknowledged.

\appendix

\section{The case of two charge carriers}
In the case of two kinds of charge carriers, for example electrons and
holes,  in the  current  two-band model, we
should solve the equations of motion (\ref{equ8}) for each kind of
charge carriers.    The total conductivity tensor
 consists of the sum of conductivity tensors of
transmission electrons and holes
\begin{eqnarray}
\sigma(\tau)=e^2
 \begin{pmatrix}
  \frac{C_3^e(\tau)n_e}{m_x^e}&+&\frac{C_3^h(\tau)n_h}{m_x^h} &&& -\frac{D_3^h(\tau)n_h}{m_y^h}+\frac{D_3^e(\tau)n_e}{m_y^e} \\
 \frac{\tilde{D}_3^h(\tau)n_h}{m_x^h} &-&\frac{\tilde{D}_3^e(\tau)n_e}{m_x^e} &&& \frac{\tilde{C}_3^e(\tau)n_e}{m_y^e}+\frac{\tilde{C}_3^h(\tau)n_h}{m_y^h}
\end{pmatrix},
\label{equa1}
\end{eqnarray}
where $n_e$ ($m_{x,y}^e$) and $n_h$ ($m_{x,y}^h$) are the concentrations (the components of the effective mass
tensor) of electrons and holes, respectively.
Performing the  inverse operation on $\sigma(\tau)$,  we  find the   magneto-resistance tensor:
\begin{eqnarray}
\rho(\tau)&=&\frac{1}{\Delta(\tau)}\times\\
 &\times& \begin{pmatrix}
 m_x^e m_x^h[ m_y^e n_h \tilde{C}_3^h(\tau)+m_y^h n_e \tilde{C}_3^e(\tau)] &&  m_x^e m_x^h[m_y^e n_h D_3^h(\tau)-m_y^h n_e D_3^e(\tau)] \nonumber\\
 - m_y^e m_y^h[m_x^e n_h \tilde{D}_3^h(\tau)-m_x^h n_e \tilde{D}_3^e(\tau)] && m_y^e m_y^h[m_x^e n_h C_3^h(\tau)+m_x^h n_e
 C_3^e(\tau)]
\end{pmatrix},
\label{equa2}
\end{eqnarray}
where
$$
\Delta(\tau)=e^2([m_x^e n_h C_3^h(\tau)+m_x^h n_e C_3^e(\tau)][m_y^e n_h
\tilde{C}_3^h(\tau)+m_y^h n_e \tilde{C}_3^e(\tau)]$$
$$+[m_y^e n_h D_3^h(\tau)-m_y^h n_e D_3^e(\tau)][m_x^e n_h \tilde{D}_3^h(\tau)-m_x^h n_e \tilde{D}_3^e(\tau)]).
$$
The Hall resistance takes   the following form
\begin{eqnarray}
\lefteqn{\rho_H(\tau)=\frac{m_x^e m_x^h[m_y^h n_e D_3^e(\tau)-m_y^e n_h
D_3^h(\tau)]}{\Delta(\tau)}}\nonumber\\
&&{}=\frac{m_y^e m_y^h[m_x^h n_e \tilde{D}_3^e(\tau)-m_x^e n_h
\tilde{D}_3^h(\tau)]}{\Delta(\tau)}.
\label{equa3}
\end{eqnarray}

\section{Variances }
The equations for the second moments (variances),
$$\Sigma_{q_iq_j}(t)=\frac{1}{2}<q_i(t)q_j(t)+q_j(t)q_i(t)>-<q_i(t)><q_j(t)>,$$
where $q_i=x,y,\pi_x$, or $\pi_y$ ($i$=1-4), are
\begin{eqnarray}
\dot{\Sigma}_{xx}(t)&=&\frac{2\Sigma_{x\pi_x}(t)}{m_x},\hspace{.3in}
\dot{\Sigma}_{yy}(t)=\frac{2\Sigma_{y\pi_y}(t)}{m_y},\nonumber\\
\dot{\Sigma}_{xy}(t)&=&\frac{\Sigma_{x\pi_y}(t)}{m_y}+\frac{\Sigma_{y\pi_x}(t)}{m_x},\nonumber\\
\dot{\Sigma}_{x\pi_y}(t)&=&-\lambda_{\pi_y}(t)\Sigma_{x\pi_y}(t)-\tilde{\omega}_{cx}(t)\Sigma_{x\pi_x}(t)+\frac{\Sigma_{\pi_x\pi_y}(t)}{m_x}+2D_{x\pi_y}(t),\nonumber\\
\dot{\Sigma}_{x\pi_x}(t)&=&-\lambda_{\pi_x}(t)\Sigma_{x\pi_x}(t)+\tilde{\omega}_{cy}(t)\Sigma_{x\pi_y}(t)+\frac{\Sigma_{\pi_x\pi_x}(t)}{m_x}+2D_{x\pi_x}(t),\nonumber\\
\dot{\Sigma}_{y\pi_x}(t)&=&-\lambda_{\pi_x}(t)\Sigma_{y\pi_x}(t)+\tilde{\omega}_{cy}(t)\Sigma_{y\pi_y}(t)+\frac{\Sigma_{\pi_x\pi_y}(t)}{m_y}+2D_{y\pi_x}(t),\nonumber\\
\dot{\Sigma}_{y\pi_y}(t)&=&-\lambda_{\pi_y}(t)\Sigma_{y\pi_y}(t)-\tilde{\omega}_{cx}(t)\Sigma_{y\pi_x}(t)+\frac{\Sigma_{\pi_y\pi_y}(t)}{m_y}+2D_{y\pi_y}(t),\nonumber\\
\dot{\Sigma}_{\pi_y\pi_y}(t)&=&-2\lambda_{\pi_y}(t)\Sigma_{\pi_y\pi_y}(t)-2\tilde{\omega}_{cx}(t)\Sigma_{\pi_x\pi_y}(t)+2D_{\pi_y\pi_y}(t),\nonumber\\
\dot{\Sigma}_{\pi_x\pi_x}(t)&=&-2\lambda_{\pi_x}(t)\Sigma_{\pi_x\pi_x}(t)+2\tilde{\omega}_{cy}(t)\Sigma_{\pi_x\pi_y}(t)+2D_{\pi_x\pi_x}(t),\nonumber\\
\dot{\Sigma}_{\pi_x\pi_y}(t)&=&-(\lambda_{\pi_x}(t)+\lambda_{\pi_y}(t))\Sigma_{\pi_x\pi_y}(t)+
\tilde{\omega}_{cy}(t)\Sigma_{\pi_y\pi_y}(t)-\tilde{\omega}_{cx}(t)\Sigma_{\pi_x\pi_x}(t)
+2D_{\pi_x\pi_y}(t).
\label{equ26C}
\end{eqnarray}
So, we   obtain  the Markovian-type (local in time) equations
for the first and second moments, but with the transport
coefficients depending explicitly on time. The time-dependent
diffusion coefficients $D_{q_iq_j}(t)$ are determined as
\begin{eqnarray}
D_{xx}(t)&=&D_{yy}(t)=D_{xy}(t)=0,\nonumber\\
D_{\pi_x\pi_x}(t)&=&\lambda_{\pi_x}(t)J_{\pi_x\pi_x}(t)-\tilde{\omega}_{cy}(t)J_{\pi_x\pi_y}(t)+\frac{1}{2}\dot{J}_{\pi_x\pi_x}(t),\nonumber\\
D_{\pi_y\pi_y}(t)&=&\lambda_{\pi_y}(t)J_{\pi_y\pi_y}(t)+\tilde{\omega}_{cx}(t)J_{\pi_x\pi_y}(t)+\frac{1}{2}\dot{J}_{\pi_y\pi_y}(t),\nonumber\\
D_{\pi_x\pi_y}(t)&=&-\frac{1}{2}\left[-(\lambda_{\pi_x}(t)+\lambda_{\pi_y}(t))J_{\pi_x\pi_y}(t)+
\tilde{\omega}_{cy}(t)J_{\pi_y\pi_y}(t)-\tilde{\omega}_{cx}(t)J_{\pi_x\pi_x}(t)-\dot{J}_{\pi_x\pi_y}(t)\right],\nonumber\\
D_{x\pi_y}(t)&=&-\frac{1}{2}\left[-\lambda_{\pi_y}(t)J_{x\pi_y}(t)-\tilde{\omega}_{cx}(t)J_{x\pi_x}(t)+\frac{J_{\pi_x\pi_y}(t)}{m_x}-\dot{J}_{x\pi_y}(t)\right],\nonumber\\
D_{y\pi_x}(t)&=&-\frac{1}{2}\left[-\lambda_{\pi_x}(t)J_{y\pi_x}(t)+\tilde{\omega}_{cy}(t)J_{y\pi_y}(t)+\frac{J_{\pi_x\pi_y}(t)}{m_y}-\dot{J}_{y\pi_x}(t)\right],\nonumber\\
D_{x\pi_x}(t)&=&-\frac{1}{2}\left[-\lambda_{\pi_x}(t)J_{x\pi_x}(t)+\tilde{\omega}_{cy}(t)J_{x\pi_y}(t)+\frac{J_{\pi_x\pi_x}(t)}{m_x}-\dot{J}_{x\pi_x}(t)\right],\nonumber\\
D_{y\pi_y}(t)&=&-\frac{1}{2}\left[-\lambda_{\pi_y}(t)J_{y\pi_y}(t)-\tilde{\omega}_{cx}(t)J_{y\pi_x}(t)+\frac{J_{\pi_y\pi_y}(t)}{m_y}-\dot{J}_{y\pi_y}(t)]\right].
\label{equ27C}
\end{eqnarray}
Here, $\dot{J}_{q_iq_j}(t)=dJ_{q_iq_j}(t) / dt$ and
\begin{eqnarray}
J_{xx}(t)&=&\ll I_x(t)I_x(t)+I'_x(t)I'_x(t)\gg,  \hspace{.1in}
J_{yy}(t)=J_{xx}(t)|_{x\to y},\nonumber\\
J_{xy}(t)&=&\ll I_x(t)I_y(t)+I'_x(t)I'_y(t)\gg,  \hspace{.1in} J_{\pi_x\pi_y}(t)=J_{xy}(t)|_{x\to \pi_x,y\to \pi_y}, \nonumber\\
J_{x\pi_x}(t)&=&\ll I_{x}(t)I_{\pi_x}(t)+I'_{x}(t)I'_{\pi_x}(t)\gg,  \hspace{.1in} J_{y\pi_x}(t)|=J_{x\pi_x}(t)|_{x\to y}, \nonumber\\
J_{x\pi_y}(t)&=&\ll I_{x}(t)I_{\pi_y}(t)+I'_{x}(t)I'_{\pi_y}(t)\gg, \hspace{.1in} J_{y\pi_y}(t)=J_{x\pi_y}(t)|_{x\to y},\nonumber\\
J_{\pi_x\pi_x}(t)&=&\ll I_{\pi_x}(t)I_{\pi_x}(t)+I'_{\pi_x}(t)I'_{\pi_x}(t)\gg,  \hspace{.1in}
J_{\pi_y\pi_y}(t)=J_{\pi_x\pi_x}(t)|_{pi_x\to \pi_y}.
%J_{y\pi_y}(t)&=&\ll I_{y}(t)I_{\pi_y}(t)+I'_{y}(t)I'_{\pi_y}(t)\gg,  \hspace{.1in}
%J_{\pi_x\pi_y}(t)=\ll I_{\pi_x}(t)I_{\pi_y}(t)+I'_{\pi_x}(t)I'_{\pi_y}(t)\gg.
\label{equ28C}
\end{eqnarray}
 The explicit expressions for   $J_{q_iq_j}(t)$
are
\begin{eqnarray}
  J_{xx}(t)&=&\frac{m \hbar \gamma^2}{\pi}\int_{0}^{\infty}d\omega\int_0^tdt'\int_0^tdt''\frac{\omega
  \coth\left[\frac{\hbar\omega}{2T}\right]}{\omega^2+\gamma^2} \nonumber\\
  &\times&\left[\lambda_xA_1(t')A_1(t'')+\lambda_yA_2(t')A_2(t'')\right]\cos(\omega[t''-t']),\nonumber
\end{eqnarray}
%\begin{eqnarray}
%  J_{yy}(t)&=&\frac{m \hbar \gamma^2}{\pi}\int_{0}^{\infty}d\omega\int_0^tdt'\int_0^tdt''\frac{\omega
%  \coth\left[\frac{\hbar\omega}{2T}\right]}{\omega^2+\gamma^2}\nonumber\\
%  &\times&\left[\lambda_xB_2(t')B_2(t'')+\lambda_yB_1(t')B_1(t'')\right]\cos(\omega[t''-t']),\nonumber
%\end{eqnarray}
\begin{eqnarray}
  J_{xy}(t)&=&\frac{m \hbar \gamma^2}{\pi}\int_{0}^{\infty}d\omega\int_0^tdt'\int_0^tdt''\frac{\omega
  \coth\left[\frac{\hbar\omega}{2T}\right]}{\omega^2+\gamma^2}\nonumber\\
  &\times&\left[\lambda_xA_1(t')B_2(t'')+\lambda_yA_2(t')B_1(t'')\right]\cos(\omega[t''-t']),\nonumber
 \label{equ29C}
\end{eqnarray}
\begin{eqnarray}
  J_{\pi_x\pi_x}(t)&=&\frac{m \hbar \gamma^2}{\pi}\int_{0}^{\infty}d\omega\int_0^tdt'\int_0^tdt''\frac{\omega
  \coth\left[\frac{\hbar\omega}{2T}\right]}{\omega^2+\gamma^2}\nonumber\\
    &\times&\left[\lambda_xC_1(t')C_1(t'')+\lambda_yC_2(t')C_2(t'')\right]\cos(\omega[t''-t']),\nonumber
\label{equ30C}
\end{eqnarray}
%\begin{eqnarray}
%  J_{\pi_y\pi_y}(t)&=&\frac{m \hbar \gamma^2}{\pi}\int_{0}^{\infty}d\omega\int_0^tdt'\int_0^tdt''\frac{\omega
%  \coth\left[\frac{\hbar\omega}{2T}\right]}{\omega^2+\gamma^2}\nonumber\\
%    &\times&\left[\lambda_xD_2(t')D_2(t'')+\lambda_yD_1(t')D_1(t'')\right]\cos(\omega[t''-t']),\nonumber
%\end{eqnarray}
\begin{eqnarray}
  J_{\pi_x\pi_y}(t)&=&\frac{m \hbar \gamma^2}{\pi}\int_{0}^{\infty}d\omega\int_0^tdt'\int_0^tdt''\frac{\omega
  \coth\left[\frac{\hbar\omega}{2T}\right]}{\omega^2+\gamma^2}\nonumber\\
    &\times&\left[\lambda_xC_1(t')D_2(t'')+\lambda_yC_2(t')D_1(t'')\right]\cos(\omega[t''-t']),\nonumber
\end{eqnarray}
\begin{eqnarray}
  J_{x\pi_x}(t)&=&\frac{m \hbar \gamma^2}{\pi}\int_{0}^{\infty}d\omega\int_0^tdt'\int_0^tdt''\frac{\omega
  \coth\left[\frac{\hbar\omega}{2T}\right]}{\omega^2+\gamma^2}\nonumber\\
    &\times&\left[\lambda_xA_1(t')C_1(t'')+\lambda_yA_2(t')C_2(t'')\right]\cos(\omega[t''-t']),\nonumber
\end{eqnarray}
%\begin{eqnarray}
%  J_{y\pi_y}(t)&=&\frac{m \hbar \gamma^2}{\pi}\int_{0}^{\infty}d\omega\int_0^tdt'\int_0^tdt''\frac{\omega
%  \coth\left[\frac{\hbar\omega}{2T}\right]}{\omega^2+\gamma^2}\nonumber\\
%    &\times&\left[\lambda_xB_2(t')D_2(t'')+\lambda_yB_1(t')D_1(t'')\right]\cos(\omega[t''-t']),\nonumber
%\end{eqnarray}
\begin{eqnarray}
  J_{x\pi_y}(t)&=&\frac{m \hbar \gamma^2}{\pi}\int_{0}^{\infty}d\omega\int_0^tdt'\int_0^tdt''\frac{\omega
  \coth\left[\frac{\hbar\omega}{2T}\right]}{\omega^2+\gamma^2}\nonumber\\
    &\times&\left[\lambda_xA_1(t')D_2(t'')+\lambda_yA_2(t')D_1(t'')\right]\cos(\omega[t''-t']).\nonumber
\end{eqnarray}
%\begin{eqnarray}
%  J_{y\pi_x}(t)&=&\frac{m \hbar \gamma^2}{\pi}\int_{0}^{\infty}d\omega\int_0^tdt'\int_0^tdt''\frac{\omega
%  \coth\left[\frac{\hbar\omega}{2T}\right]}{\omega^2+\gamma^2}\nonumber\\
%    &\times&\left[\lambda_xB_2(t')C_1(t'')+\lambda_yB_1(t')C_2(t'')\right]\cos(\omega[t''-t']).\nonumber
%\end{eqnarray}

In our treatment
$D_{xx}=D_{yy}=D_{xy}=0$, because there are no random
forces for   $x$ and $y$ coordinates in Eqs.~(\ref{equ8}). At
$\omega_{cx}=\omega_{cy}=0$,  we have $D_{y \pi_x}(t)=D_{x\pi_y}(t)=D_{\pi_x \pi_y}(t)=0$.

\end{document}